\documentstyle[12pt,aaspp4]{article}

\begin{document}

\title{Multiple object and integral field near-infrared spectroscopy
using fibers}

\author{
Roger Haynes\footnotemark[1],
David Lee\footnotemark[2],
Jeremy Allington-Smith,  
Robert Content, 
George Dodsworth,
Ian Lewis\footnotemark[2],
Ray Sharples,
James Turner,
John Webster, 
Christine Done, 
Reynier Peletier
}

\affil{Astronomical Instrumentation Group, University of Durham,\\
South Rd, Durham DH1 3LE, UK}

\author{Ian Parry}
\affil{Institute of Astronomy, University of Cambridge,\\
Madingley Rd, Cambridge, CB3 0HA, UK}

\author{Scott Chapman}
\affil{Dominion Astrophysical Observatory, Herzberg Institute of
Astrophysics\\ 
5071 W. Saanich Rd, Victoria, BC, V8X 4M6, Canada}

\footnotetext[1]{E-mail: roger.haynes@durham.ac.uk}
\footnotetext[2]{Now at the Anglo-Australian Observatory}

\begin{abstract}

We describe a new system for multiple object spectroscopy and integral field
spectroscopy at near-infrared wavelengths using optical fibers. Both modes
of the SMIRFS instrument have been tested at the UK Infrared Telescope with
the CGS4 infrared spectrograph. The modular system includes a common optical
system to image the fiber slit onto the cold slit inside the CGS4 cryostat.
The multiobject mode consists of 14 interchangeable fused silica or
zirconium fluoride fibers each with a field of 4 arcsec.  The integral field
mode consists of 72 fused silica fibers coupled with a lenslet array to give
a contiguous field of 6$\times$4 arcsec with 0.6 arcsec sampling.

We describe the performance of both modes. For the multiobject mode, the
feasibility and desirability of using fluoride fibers to extend the
wavelength range into the K-band is discussed. For the integral field mode,
the performance is compared with theoretical expectation with particular
attention to the effect of Focal Ratio Degradation in the fibers.

These results demonstrate the feasibility of multiobject and integral field
spectroscopy in the near-infrared using lenslet-coupled fiber systems.
Although SMIRFS in an experimental system working with a spectrograph not
designed for this purpose, the throughput and uniformity of response are
good. SMIRFS points the way forward to systems with much larger numbers of
elements.

\end{abstract}


\section{Introduction}

SMIRFS is an instrument to explore new techniques for multiobject
spectroscopy (MOS) and integral field spectroscopy (IFS) at near-infrared
wavelength (1-2.5$\mu$m).  It was developed at low cost to provide the UK
Infrared Telescope (UKIRT) with a simple capability in these areas and to
develop the techniques needed to build larger-scale instruments for 4-m and
8-m telescopes. As such, the instrument was designed to a modest
specification with the aim of building it quickly and obtaining useful
results of use to other projects.  The MOS and IFS modes of SMIRFS are
referred to as SMIRFS-MOS and SMIRFS-IFS respectively.

In section 2, we describe the basic SMIRFS system consisting of the optical
relay from the fiber slit to the cryogenic slit inside CGS4 and the
multi-fiber system of the MOS mode. In Section 3, we describe the
performance of this mode and present an example dataset obtained during
commissioning. The IFS mode is described in Section 4. Its performance is
presented in Section 5 which includes a comparison with theoretical
expectations. In section 6, we give an example of a dataset obtained during
commissioning which also serves to illustrate the operation of an integral
field spectrograph. In Section 7, we state our conclusions and discuss the
relevance of this work to other IFS systems under construction. We start by
discussing the motivation for this work.

\subsection{Multiobject spectroscopy in the infrared}

To date, the technique of multiple object spectroscopy (MOS) has been almost
exclusively employed at visible wavelengths.  Although {\it Multi-slit}
spectroscopy (e.g.  LDSS-2, Allington-Smith et al. 1994\markcite{jras94},
and GMOS, Murowinski et al.  1998\markcite{murowinski98}) provides the best
background subtraction because contiguous regions of sky within the same
slit are sampled, this is bought at the expense of truncation of spectra
near the edge of the field and some problems in addressing real target
distributions.  In contrast {\it multi-fiber} systems
(e.g. Autofib-2/WYFFOS, Parry et al.  1994\markcite{parry94}, and
2dF\markcite{Taylor97}, Taylor 1997) avoid these problems because the
spectrum layout on the detector is independent of the field. However the
accuracy of background subtraction is generally worse leading to limiting
magnitudes which are 1-2 magnitudes brighter (see also Cuby \& Mignoli
1994\markcite{cuby94}) and there are problems in addressing dense target
distributions due to fiber collision restrictions.

The argument for Multiobject spectroscopy is just as strong in the infrared
as in the visible, especially since the spectral energy distributions of
field galaxies increasingly redshift to longer wavelengths so that the most
useful spectral features appear in the near infrared.  As an example of the
surface densities to be encountered, the field galaxy population contains
$\sim 10^4$/deg$^2$ at $K=20$ which gives a multiplex gain of 10 in fields
as small as 2 arcmin.

Extending multiobject techniques into the infrared is not straightforward.
The difficulty of building a reconfigurable multiobject capability into a
cryogenic instrument has so far ruled out signal/noise-optimized MOS at
longer wavelengths ($> 1.8 \mu$m) where it is necessary to encapsulate the
optical system in a cryostat, although various systems are under development
(e.g. EMIR for Gran Telescopio Canarias and the GIRMOS concept for GEMINI).
However at shorter wavelengths, where the instrumental thermal background is
not a problem ($1-1.8\mu$m), uncooled multi-fiber and multi-slit methods may
be used
(Herbst et al. 1995\markcite{herbst95}).
Another issue is that current infrared instruments employ smaller detectors
than in the visible. As we have seen, this poses some difficulties for
multislit systems since they do not make optimum use of the detector surface
whereas fibers can use this area more efficiently. This aspect makes it
easier for multi-fiber systems to reach the high spectral resolutions
required to reject the strong atmospheric OH emission lines which form most
of the background in the J and H bands.

Consequently we decided to explore the use of a fiber-based MOS system in
the near-IR up to the K-band. Since this wavelength range extends into the
region where some drop in the performance of fused silica (FS) fibers may be
expected ($> 2\mu$m), we decided to provide an alternative zirconium
fluoride (ZF) fiber system.  The use of ZF was also motivated by the
possibility that fiber systems could be used in cryogenic instruments in
which case fluoride fibers would definitely be required at the longer
wavelengths which would then be accessible (up to 5$\mu$m).

For reasons of cost, it was decided to use this prototype system with an
existing infrared spectrograph: CGS4\markcite{ramsay94} (Ramsay-Howatt 1994)
on UKIRT.  This is a cryogenic instrument so a means had to be found to
inject the reformatted light into it without rebuilding the cryostat. This
was done using an optical relay between the uncooled fiber slit and the cold
spectrograph slit which fits into the space normally occupied by CGS4's
calibration unit. The system was commissioned in June 1995 and December
1996. The first run was affected by very poor conditions. The second run had
slightly better luck and incorporated an optional output mask to cut down
thermal background from the fiber slit. Thereafter the system was adapted
for integral field spectroscopy as described below.

\subsection{Integral field spectroscopy in the infrared}

Integral field spectroscopy (IFS) produces a spectrum from each part of a
two-dimensional field (e.g. Bacon et al.\markcite{bacon95} 1995).  In
contrast, long-slit spectroscopy is limited to a one-dimensional field whose
width is determined by the need to obtain good spectral resolution.  IFS
avoids this restriction by decoupling the slit width from the field shape by
reformatting a rectangular field into a linear pseudo-slit. A further
advantage is that precise target acquisition is not required since the
object does not need to be carefully placed on a narrow slit. If desired,
the acquisition can be checked by reconstructing a white-light image of the
object by summing the two-dimensional spectrogram over wavelength.  Even
when observing unresolved objects in poor seeing, the system acts as an
image slicer to eliminate slit losses.

The scientific motivation for integral field spectroscopy is summarized in
Allington-Smith \& Content (1998\markcite{ac98}; hereafter AC) which also
includes a discussion of sampling and background subtraction issues relevant
to fiber-lenslet integral field spectrographs and describes the basic
techniques.  The fiber-lenslet technique used in SMIRFS ensures that
the field is contiguous, with unit filling factor, whilst maximising
throughput by optimal coupling with both the telescope and spectrograph.

Particular applications for IFS in the infrared include studies of the
obscured nuclear regions of active galaxies; the optical-radio co-alignment
of distant radio galaxies, and studies of shocks in star forming regions. An
example given in this paper relates to active galaxies via imaging of
diagnostic emission lines indicative of star formation and non-thermal
emission. Many of the key diagnostic spectral features for star-forming
regions appear in the infrared and familiar ionic emission lines in distant
galaxies are redshifted into the infrared.

As discussed by AC, the fiber-lenslet technique provides significant
benefits over lenslet-only systems (e.g. TIGER, Bacon {\it et
al.}\markcite{bacon95} 1995) in terms of the efficiency with which the
detector surface is addressed and the length of spectrum which can be
obtained without overlaps between spatial elements. A lenslet-only system
for UKIRT, where the available number of pixels was initially only $256
\times 72$ pixels, would involve significant compromise in field and
spectrum length.  The fiber-lenslet approach also has advantages over
fiber-only systems.  Firstly, it provides much better coupling to the
telescope and spectrograph. Without this, the very slow beam from the
telescope (f/36) would result in very low throughput.  Secondly, it provides
unit filling factor whereas a fiber-only system wastes the light which
strikes the cladding (and buffer, if present) between fiber cores.
Therefore, the fiber-lenslet approach is the best for implementing an IFS
capability on UKIRT.  This approach has also been adopted by us for the
Thousand-element integral field unit (TEIFU; Haynes et al.
1998\markcite{haynes98}), for the IFS capabilities of the VLT VIMOS (Lefevre
et al. 1998\markcite{lefevre98}) and for the GEMINI Multiobject
Spectrographs (GMOS, Allington-Smith et al.
1997\markcite{jras97}). Other systems using this technique are SPIRAL and
COHSI (Kenworthy et al. 1998\markcite{kenworthy98}).

For these reasons, we decided to adapt the SMIRFS-MOS system to a
fiber-lenslet integral field unit (IFU).  This system, which re-uses the
SMIRFS-MOS infrastructure, allows us to prove technology to be applied to
TEIFU (now successfully commissioned) and the GMOS IFU. 

The results in this paper refer to two observing runs with the SMIRFS-IFU
system.  One in June 1997, for initial technical commissioning was
accompanied by very poor weather but the other, in March 1998, was more
successful and allowed spectral-line mapping of active galaxies (Turner {\it
et al.} in preparation, Chapman et al. in preparation). The first run
used CGS4 with its short camera giving a sampling of one detector pixel per
spatial element while the second run used the long camera with a sampling of
2 pixels per spatial element.

\section{The SMIRFS-MOS system}
 
The system is described in detail by Haynes (1995)\markcite{haynes95}.
Fig.~\ref{fig-plan} shows the layout of SMIRFS and its coupling to the
telescope and CGS4.  SMIRFS has been designed to mount onto the West Port of
UKIRT, which is reserved for visiting instruments. There are essentially
four parts to the system; the field plate unit, the guide fiber unit, the
fiber bundle and the slit projection unit.  Despite the name, the guide
fiber unit is not usually used for guiding, but is mainly used for field
acquisition and checking the field plate orientation.

\subsection{Field plate unit}

The field plate unit's function is to hold the field plate and thereby the
input of the fiber bundle at the focal plane of the telescope. The focus of
UKIRT is 196mm from the West face of the instrument support unit. The field
plate contains pre-drilled holes which correspond to the positions of the
objects to be observed as well as extra holes to hold the acquisition fibers
and dedicated sky fibers, if required. A different field plate is needed for
each field. The fibers are fixed in small brass ferrules which are held in
the field plate by a lock nut (Fig.~\ref{fig-ferrule}). The field may be
adjusted in rotation to correct for any misalignment between the field plate
and the object field. The correction is performed once during the instrument
set-up and does not normally require changing when a different field plate
is installed. However, as a quick check, the guide fibers can be used to
ensure the orientation is correct. The field plate unit also has a plate
tensioner, that pulls on the centre of the field plate to distort it to
approximate a spherical surface with a radius of 11.5m to ensure that all
the fibers in the field are pointing correctly at the exit pupil of the
telescope (the secondary mirror). Any error in the global fiber pointing can
be corrected using the UKIRT dichroic mirror, which directs the infrared
light to the required port, while letting the visible light through to the
cross-head acquisition and guide camera. The available field is 4 arcmin
across and the minimum fiber to fiber spacing is approximately 18
arcsec. Fibers may not be deployed within the central 14 arcsec, where the
tensioner is attached.

\subsection{Guide fiber unit}

This unit consists of three fiber bundles. Each is a coherent bundle
containing 7 fibers; one central fiber closely surrounded by a ring of
6. These are coupled to a camera via two magnifying lenses. This enables the
operator to determine the centroid of up to three stars in the field and
from that make corrections to the telescope pointing and field plate
rotation. Originally a SCANCO intensified camera was used, but this was not
sufficiently sensitive to the small amount of predominately red light that
was reflected by the dichroic. It was later replaced with a CCD, which
considerable improved the sensitivity. However, because the guide fiber unit
uses the visible light that is reflected by the dichroic, a correction for
atmospheric refraction may be required at large zenith distance. The
diameter of a single guide fiber bundle corresponds to approximately 2.7
arcsec on the sky.

\subsection{Multi-object fiber bundles}

In MOS mode, SMIRFS has two fiber bundles each containing 14 fibers: a
zirconium fluoride (ZF) system for use in the K band and an ultra-low OH
fused silica (FS) system for the J and H bands.
The throughput of different fiber type is shown in Fig~\ref{fig-fthro}.
A small CaFl$_2$ lenslet (Fig.~\ref{fig-ferrule}) is used at the input and
output of each fiber to couple the telescope beam (f/36) into the fiber at
f/5 and then back to f/36 at the output for coupling to the
spectrograph. This reduces losses due to focal ratio degradation\footnote{A
non-conservation of Etendue which results in the output beam being faster
than the input beam. It is an equivalent to an increase in entropy so should
be avoided.}  (FRD) within the fibers.  The input lenslet (diameter 3mm,
focal length 7.6mm) re-images the telescope exit pupil onto the face of the
fiber core (200$\mu$m diameter). The pupil size was chosen to almost
completely fill the fiber core so as to reduce the amount of light
contamination from the sky around the secondary mirror, which is unbaffled
at UKIRT. This also reduces thermal contamination from any telescope
structure around the top end.

At the bundle output, the fibers are re-formated into a long slit.  Each
fiber is coupled to the CGS4 spectrograph via another lenslet which is
identical to the lenslet at the fiber input.  The centre to centre spacing
along the fiber slit is 4mm, which corresponds to 6.2 arcsec at the
detector. This corresponds to 5 pixels with the short camera (focal length
150mm) and 10 with the long camera (focal length 300mm). The fiber output
spacing was constrained primarily by the number of detector pixels that are
available along the CGS4 slit and the desire to maximize the multiplex gain.
SMIRFS was original designed for use with the 256$\times$256 array and the
short camera which results in only 72 pixels along the slit.

The lenslets limit the aperture viewed by the fiber to 4.5 arcsec. However,
this can be reduced to 2 arcsec in good seeing conditions by the use of a
field stop mounted on the front of the fiber ferrule. This is not advisable
for K band spectroscopy as the stop will contribute significantly to the
thermal background as it is not cooled.

The thermal background is kept to a minimum by using the CGS4 spectrograph's
cooled Lyot stop to mask out any contamination from beam angles faster than
f/36 and by imaging the fiber slit onto the cooled CGS4 long slit. After the
first telescope run it was found that there was a significant amount of
thermal background from the slit material between the fibers. The slit was
then modified by placing a reflective mask just in front of the fiber
slit. This contained 14 holes to allow the light from the fibers to pass
through and was positioned at such an angle that only light which had
originated in CGS4 was reflected back into it, which, being cooled, produces
very little thermal background.  Each hole acts as a field stop limiting the
effective aperture to $\sim$2.5 arcsec on the sky. However, the aperture is
slightly blurred by FRD effects so some light will come from outside a
diameter of 2.5 arcsec. The slit mask considerably reduced the thermal
contamination from the slit material, and is discussed in the next section.

\subsection{Slit projection unit}

As the spectrograph is a cryogenic instrument it was impractical to replace
the CGS4 long slit with the SMIRFS fiber slit. It was therefore necessary to
project the image of the fiber slit through the cryostat window into
CGS4. This is achieved using a spherical re-imaging mirror and a plane fold
mirror. In order to imitate the flat slit of CGS4 it was necessary to curve
the fiber slit to match the field curvature of the re-imaging mirror, thus
producing an image magnification of unity. Both mirrors are mounted on a
tip/tilt and translation stage to facilitate alignment of the fiber slit
with the CGS4 optical axis.

\section{SMIRFS-MOS performance}

\subsection{Efficiency}

The first commissioning run took place in June 1995 (Haynes
1995\markcite{haynes95}, Haynes et al. 1995\markcite{haynesetal95}) using
the short CGS4 camera and a 75 lines/mm grating. The run was badly affected
by poor weather but the provisional results obtained led to a number of
modifications, the most significant of which was the reflective mask for the
ZF fiber bundle to reduce the thermal background in the K band.

The second run in December 1996 used the same spectrograph configuration as
June 95. The weather was partially clear but unsuitable for photometric
studies much of the time. During the remainder of the time, photometric
variation was around 10\% which limits the accuracy of the results
presented. The purpose of the run was primarily to observe a number of K and
M giant star in both open and globular clusters to establish a new method of
metal abundance determination for cool stellar population and demonstrate
the potential of SMIRFS and future infrared fiber systems.

A summary of the throughput performance of SMIRFS (CGS4 short camera and
grating) is given in Table~\ref{tab-moseff}. This gives the efficiency of
the SMIRFS alone, obtained from a comparison of the count rates with the
SMIRFS-MOS installed and with it removed from from CGS4 so that light enters
the spectrograph directly through the slit.

The predicted values are estimates that take into account the average fiber
transmission, reflection losses and optical alignment errors (see Haynes
1995\markcite{haynes95} for further details). The FS fiber throughput agrees
reasonable well with the predictions, especially in the H and K bands. The
results for the ZF without the output mask bundle agree well with the
predictions, but the results with the output mask installed are
significantly lower. This may be due to vignetting by the mask, resulting
from mis-alignment of the star on the fiber input or vignetting by the CGS4
slit due to flexure (Kerr 1997\markcite{kerr97}).

The problem of flexure within CGS4 became more apparent during the
SMIRFS IFU run discussed later. The fiber to fiber throughput variations
are $\sim$10\% RMS (Fig.~\ref{fig-mosflat}) and are dominated by
errors in the alignment between the fiber inputs and the telescope
pupil.

\subsection{Image quality}

The PSF of the system can be demonstrated from a comparison of an
observation of a standard star with SMIRFS and by CGS4 alone using just
a long slit (Fig.~\ref{fig-mospsf}). Gaussian fits to the cores of the
two profiles indicate $\sigma=0.56$ and $0.64$ pixels with and without
SMIRFS respectively. Although this comparison does not take the seeing
into account, it suggests that any degradation in spatial resolution by
SMIRFS is small and that the wings of the distribution arise from within
CGS4. This is not surprising since the radial distribution of light in
the object should be preserved at the slit (because the fibers preserve
the radial {\it angular} distribution of light), except for the effect
of FRD.  This suggests that FRD is not a significant problem.  No
obvious sign of truncation of the PSF due to the effect of the output
mask is seen: the results are similar for the FS (unmasked) and ZF
fibers (masked and unmasked).

\subsection{Background removal}

Although background subtraction, by e.g. beam-switching, will remove the
background whether it arises from the instrument or the sky, the
signal/noise is degraded by the photon noise from the background. Therefore
it is important to minimize the source of background before it is recorded
by the detector.

Although the poor conditions during the June 1995 run mean that it is not
possible to make a quantitative comparison of the thermal background of
SMIRFS with and without the output slit mask, a large reduction in the
inter-fiber background for the K band was noted after the output mask was
installed. When observing sky the signal from the inter-fiber area was less
than the signal from the fiber across the whole of the K band whereas
previously it was often higher. This is shown in Fig.~\ref{fig-back} where
the background count rate using the mask is plotted separately as the
component from the fiber alone and from the inter-fiber gap. This excludes
thermal background from the sky and telescope but includes any contribution
from CGS4 since this could not be determined independently. For this reason
the count rates are upper limits.

Before the output mask was installed, a comparison of data taken with and
without SMIRFS showed that the thermal background between 2.2 and 2.5$\mu$m
was only $\sim$3 times larger for SMIRFS than for CGS4 alone (Haynes
1995\markcite{haynes95}).  Although we were unable to make a direct
comparison after the mask was installed, we believe that the thermal
background due to SMIRFS is significantly less than implied by the figures
given above.

\subsection{Examples of data}

An example of the K band spectra from giant stars in a globular cluster
field (NGC1904) is shown in Figs~\ref{fig-image} and \ref{fig-spectra}. Beam
switching was used for sky subtraction, by means of a telescope nod,
typically a few arcmin. The observing sequence was 12 exposures of 5 seconds
with $2\times2$ sampling (4 minutes) in a series of on-off-off-on target
positions. This was repeated 8 times to give a total on-target time of 64
minutes. The data has been wavelength calibrated with an Argon arc spectrum
and then co-added. The typical NaI (2.204$\mu$m) and weak CaI (2.258$\mu$m)
absorption, plus the CO bands at redder wavelength (Terndrup et
al. 1991\markcite{terndrup91}), are visible in the brightest of the
objects. It is also encouraging that the fibers dedicated to sky indicate
negligible sky subtraction errors.  Table~\ref{tab-mosobs} contains a list
of the objects, their magnitudes and broad-band colors in the top-bottom
order in which they appear in Fig.~\ref{fig-spectra}.

\section{The SMIRFS-IFS system}

\subsection{The principle of fiber-lenslet IFS systems}

As described in Section 1.2, fiber-lenslet systems have advantages over both
lenslet-only systems in terms of the efficiency with which the detector
surface is used --- leading to an increase in field of view for fixed
sampling increment and detector format --- and fiber-only systems in terms
of efficiency and filling factor.  The basic principle of the SMIRFS
integral field unit (IFU) is shown in Fig.~\ref{ifu-scheme}. An image of the
sky is formed on the input lenslet array. This forms images of the telescope
pupil on the cores of a matching array of fibers bonded to the unfigured
side of the lenslet array. This performs two functions: firstly, the beam is
made faster to reduce the effect of FRD (Section 2.3) and, secondly, all the
light falling on the lenslet aperture is captured by the fiber leading to
almost unit filling factor. It is very important that as much light as
possible is directed into the fibers which requires careful control of
positional errors between the lenslets and fibers. In a visible-light system
the fiber cores could be oversized but this is not desirable in the infrared
because it would inject background light into the fibers. For this reason,
the core size is well-matched to the telescope pupil.  The fibers are
reformatted into a pseudoslit which consists of a line of fibers bonded to a
linear lenslet array. The output of each fiber is a scrambled image of the
telescope pupil from which each lenslet forms a scrambled sky image while
ensuring that the beam is well-matched to CGS4.  The resulting line of
images, which forms the actual pseudoslit, is then re-imaged onto the cold
slit of CGS4 inside its cryostat in the same was as for the SMIRFS-MOS
system.

Overlaps in the distribution of light between elements at the slit are
permissible since this only results in a small degradation in spatial
resolution in the slit direction (AC). But this requires that the mapping
between the IFU input and output is such that objects which are adjacent on
the sky are also adjacent at the slit. This condition is satisfied by the
`snakewise' mapping shown in Fig~\ref{ifu-input}. To completely avoid
overlaps would require a separation between the IFU outputs at the slit
which would drastically reduce the number of spatial elements.

\subsection{Description of the SMIRFS IFU}

Full details can be found in Lee (1998).\markcite{lee-thesis}
Table~\ref{tab-ifu} provides a summary.  The SMIRFS-IFU reformats a
rectangular field of 6$\times$4 arcsec onto the CGS4 slit. The field is
sampled by 72 hexagonal elements of size 0.6 arcsec. A spectrum is produced
from each element when dispersed by the spectrograph. The number of elements
is determined by the slit length of CGS4. With the short camera, the
sampling is 1 pixel/element, but 2 pixels/element if the long camera is
used.  The sampling scale was a balance between the desire to exploit good
images available from the UKIRT tip-tilt secondary mirror and the need to
cover a useful field of view, given the limited number of elements.

The system consists of a field plate unit which supports the input of the
IFU at the telescope focal plane, the lensed fiber bundle and a slit
projection unit that projects the IFU slit onto the CGS4 slit. The slit
projection unit and field plate unit are part of the SMIRFS-MOS system.  The
system is uncooled since it is not possible to package the system within the
CGS4 cryostat. For this reason, the system is optimized for the J and H
bands, although some useful performance may be expected in the K band,
albeit with some instrumental thermal background.

The field plate unit is mounted on a port of the Instrument Selection Unit
at 90$^\circ$ azimuth to CGS4. When the dichroic is correctly positioned,
infrared light from the sky is focussed onto the IFU input. Visible light
continues through the dichroic and forms an image on the cross-head camera
which is used for acquisition and guidance.

The IFU input (Fig~\ref{ifu-input}) consists of a microlens array which
forms images of the telescope pupil on the fiber cores.  The fibers
(Table~\ref{tab-fibers}) are individually located in an array of microtubes
(Fig. \ref{ifu-tubes}) to an accuracy of 7$\mu$m RMS.  The overall RMS
positional accuracy of the fiber cores with respect to the lenslets is
$\leq10\mu$m. This includes the non-concentricity of the fibers with respect
to their outer diameter, the non-concentricity of the fibers with the
microtubes and positional errors within the microlens array.  A small
chimney baffle reduces background contamination by preventing light from
entering the fibers at large angles.

The fibers are grouped together in a conduit and led into the slit plate
unit which is installed in place of the CGS4 calibration unit. The fibers
are terminated at 12 {\it slit blocks} where the fibers are interfaced to
linear microlens arrays to form the slit (Figs~\ref{ifu-slit} and
\ref{ifu-slitblock}).  The fibers were found to be positioned with an
accuracy of 7.5$\mu$m RMS with the largest error being 23$\mu$m. The fibers
are attached to the flat end of the output microlens arrays
(Fig.~\ref{ifu-outlens}) which focus parallel rays emerging from the fibers
to form a scrambled image of the sky within each of the subapertures at the
location of the IFU slit.  Each block is angled to ensure that the principal
arrays arrive at CGS4 with the same angle as when beam-fed directly from the
telescope.

The lenslet arrays were made by {\it Adaptive Optics Associates} and consist
of convex lens surfaces replicated in epoxy on a fused silica substrate.
Tests of the lenslets show good position accuracy within the
array. Measurements of the PSF for both input and output arrays show a
well-concentrated diffraction-limited core with extended wings (see Section
6.4.3 of Lee 1998\markcite{lee-thesis} and Lee et al.
1998\markcite{lee98}).  The wings are believed to arise from small defects
in the lens surfaces.
Table~\ref{tab-ifseff} shows detailed predictions from a numerical model of
the IFU efficiency using realistic models of the loss mechanisms based on
measurements of the lenslet PSF and FRD taking diffraction into account and
assuming that the scattering losses
are proportional to the inverse square of wavelength (Nussbaum et al.
1997\markcite{nussbaum}).  This suggests that the dominant loss mechanism
($\sim$20\%) is the mismatch between the fiber core and the imperfect pupil
image produced at the fiber input. Recall that the pupil image was not
undersized with respect to the fiber core because of the need to baffle
light from outside the telescope exit pupil.

The throughputs of the epoxy used in the lenslet fabrication and the
adhesive used for bonding to the fiber array are shown in Fig
\ref{fig-epoxyglue}. Note that the lenslet arrays are not antireflection
coated as the epoxy surface is not suitable for this process.

The fiber slit is reimaged onto the cold CGS4 slit using a spherical mirror
and a flat mirror mounted in the slit plate unit. By adjustments of these
mirrors the magnification, offset and focus of the image of the IFU slit can
be adjusted. CGS4 is used with a 2-pixel wide slit to cut down scattered
light while not vignetting the IFU slit, whose elements each project onto a
circle of diameter $\sim$1 pixel.

In the construction, much care was devoted to the following critical
areas:

\begin{enumerate}

\item Random lenslet-fiber misalignments. As discussed above, these are
such that the RMS misalignment at input and output is no greater
than 10$\mu$m RMS.

\item The fibers were polished as complete units within their respective
holders to ensure a flat surface for mating to the lenslet arrays. An
additional advantage of using lenslets is that any residual roughness in
the fiber ends is filled in by the optical adhesive.

\item FRD. This was measured for the completed fiber bundle before the
lenslet arrays were bonded and found to be such that a parallel beam at
the input produced a cone of light corresponding to $\sim$f/10 at the
output.
This was achieved by careful adjustment of the fiber bundle conduit to
relieve stress on the fibers.

\item Alignment of the lenslets with the fibers. This is a complex procedure
which eliminates not only global shifts and rotation of the lenslet arrays
with respect to the fibers but also the non-telecentricity of the telescope
feed to the spectrograph to ensure correct pupil alignment. It was estimated
that the angular misalignment about the optical axis was $<0.3^\circ$ and
the positional accuracy was $<20\mu$m. Any global shift is compensated for
during alignment at the telescope.

\end{enumerate}

\section{On-telescope performance} 

\subsection{Alignment and flexure}

At the telescope, the following alignments were carried out.

\begin{enumerate}

\item The IFU input was aligned so that the telescope exit pupil (the
secondary mirror) was aligned with the fiber cores. This was done by
back-projecting light from the fiber slit onto the telescope secondary
mirror and adjusting the angle of the dichroic which diverts infrared
light to
the IFU.

\item The relay-optics were adjusted to give unit magnification and to
eliminate angular misalignment between the fiber slit and cold 
slit. This was achieved by masking the input lenslet array to produce a
diagnostic pattern at the slit which was recorded by the science
detector.

\item Flexure test were done to examine the structural stability of the IFU
plus spectrograph. Some flexure was found amounting to a maximum of 1.1
pixels (Lee 1998\markcite{lee-thesis}). However this is believed to be
caused by flexure within CGS4 since it is similar to recent measurements
made on CGS4 alone (Kerr 1997\markcite{kerr97}).

\end{enumerate}

\subsection{Efficiency}

The efficiency of the IFU was measured by comparing observations of standard
stars taken with CGS4 alone and with the IFU plus CGS4 with a wide slit.
The results are summarized in Table~\ref{tab-ifseff}. This excludes the
efficiency of CGS4, its detector, the telescope and atmosphere.  Considering
that this instrument is a retrofit to a spectrograph not designed for IFS,
the relative throughput of $\sim$50\% is very good.

One uncertainty in this estimate is whether all the light from the star
passed through the CSG4 slit when used without the IFU. This can be
estimated from the spatial profile along the slit on the assumption that the
profile in the perpendicular direction is the same. Unfortunately, it is not
known if the extended wings of the spatial profile (see
Fig.~\ref{fig-mospsf} for an example of an observation with CGS4 alone)
arise from within the spectrograph (as suggested in Section 3.2) or before
the slit. If the latter is true, the measured throughput values should be
multiplied by 0.92. The theoretical efficiency prediction in the table
includes all known sources of error including misalignment errors, FRD and
lenslet scattering based on laboratory measurements.

\subsection{Uniformity and background subtraction}

An important aspect of fiber-lenslet IFUs is the uniformity of response over
the extent of the field. Large variations in response have the potential to
complicate data reduction and, even if they can be removed by flatfield
calibration, will compromise the final signal/noise since results will be
degraded by regions of low response. The uniformity of response is indicated
in Fig.~\ref{ifu-flat}.  There is significant variation with an RMS of
16\% of the mean.
This includes one fiber which was broken during manufacture and others where
it was noted during manufacture that the fiber was significantly displaced
from its correct position.  Nevertheless, these variation can be removed
using standard flatfield calibration to within the limits imposed by photon
statistics (Lee 1998\markcite{lee-thesis}).

Since there is no field dedicated to background subtraction, because of the
limited slit length of CGS4, the technique of beam-switching is used. This
is the same technique routinely used for CGS4 longslit observations. The
telescope is nodded between two positions on (A) and off (B) the target in
the sequence ABBA which is repeated as often as required. If the exposure
time per pointing is short enough the background will be accurately
subtracted. The temporal power spectrum of background variations is a matter
of some controversy and varies from night to night and site to site. Our
data show background residuals consistent with the results expected for CGS4
alone. There is no reason to believe that the IFU is more susceptible to
these problems than any other instrument using this technique. However,
ideally the IFU should be equipped with a dedicated field for background
subtraction so that background estimates can be obtained contemporaneously
(AC).

\subsection{Point spread function}

Here we assess the point-spread function (PSF) due to single elements of the
IFU as a test of the quality of the system. The actual PSF observed will
depend on the seeing and the method of reducing the IFS data. It is
important to distinguish the PSF measured in the {\it slit direction} and
{\it dispersion direction} in the raw data from the {\it spatial} PSF
measured in orthogonal directions in the reconstructed image of the field.

The PSF in the spectral direction was measured from observations of
wavelength calibration sources. CGS4 has the option to dither the detector
(moving the detector by sub-pixel amounts before re-combining into a single
frame in software) which allows the PSF to be properly sampled. In the
J-band, the recorded line FWHM was equivalent to using a 1.1-pixel wide
slit. In the H-band the effective IFU slit width is 1.2 pixels.  With the
IFU, CGS4 is normally used with a 2-pixel wide slit to ensure that a large
fraction of the light from the fiber slit passes into the spectrograph. In
fact, observations with CGS4 with a 1-pixel wide slit gives
FWHM$=1.15\pm0.05$ pixels which implies that use of the IFU does not degrade
the spectral resolution. However, the spectral PSF does vary slightly from
fiber to fiber in the sense that fibers with lower efficiency also produce
broader profiles (by up to $\sim$20\%). This may be because these fibers are
affected by FRD for which light exits from the fiber core at larger angles
than usual leading to a broadened PSF at the slit. This would produce both a
broadened image on the detector and lower efficiency since light emerging
from the fibers at large angles would be vignetted by the output
microlenses.

Checks were also done to see if the images at the slit were displaced in the
dispersion direction. This would complicate the interpretation of data
because of overlaps between images from adjacent fibers. This test indicates
that while there was a small non-linear distortion term amounting to
$\sim$0.1 pixel from a linear fit, no fiber output was displaced by more
than 80$\mu$m from the average position.

Finally it should be noted that the reconstructed image profiles are
slightly broadened in the direction which corresponds to the slit direction
because of the overlap of images at the slit. As discussed by AC, this
results in a small degradation in spatial resolution in this direction but
has little effect on the spatial resolution in the orthogonal
For the image shown in Fig.~\ref{fig:specmap}, which is mostly unresolved,
the cores of the profiles are well-fit by gaussian functions with FWHM of
1.32 and 1.07 arcsec in the direction parallel and perpendicular to the slit
respectively. This is consistent with a seeing FWHM of 1.0 arcsec which
indicates that the object is undersampled.  AC provided a methodology for
estimating this broadening effect which is due to the overlaps between
images at the slit and depends on the amount of FRD present.  However, in
practice, the broadening is likely to be dominated by uncompensated flexure
(Section 5.1)
which will also produce a preferential broadening in the slit direction, and
which may be present at the level of a few tenths of an arcsec.

\section{Observing with the IFU}

Here we outline the procedures required for data reduction. We present the
example of an observation of NGC4151 to illustrate the operational
techniques required.

\subsection{Data reduction}

Dispersed images from the CGS4 spectrograph are captured by a $256\times
256$ InSb detector. With the SMIRFS-IFU and long camera, a $256\times 178$
subsection is read out non-destructively, effectively providing
bias-subtracted images by comparing consecutive exposures. A single
integration, limited in duration mainly by background variability, may
consist of a number of exposures which are individually short enough to
avoid saturation. These are automatically combined by control software to
form the raw data available to the user.

The raw integrations are typically obscured by high levels of dark current
in $\sim 4\%$ of pixels, but this is removed effectively by sky
subtraction. For extragalactic work, noise is sky dominated (or dominated by
dark current for some pixels). To achieve full spectral resolution, critical
sampling of the slit width requires a shift of the detector by half a pixel
between integrations and interlacing pairs of images. In practice extra
steps are used, covering two pixel widths, to reduce the impact of dead
pixels.

The process of reconstructing three-dimensional ($x,y,\lambda$) maps of
target objects from the raw CGS4 integrations can be broken into the
following main steps:

\begin{enumerate}

  \item For each object integration (Fig.~\ref{fig:rawspec}), the
        corresponding sky integration is subtracted, removing background
        and dark current.

  \item The integration is divided by a detector flat image, taken
        before installing the IFU.

  \item Integrations taken at different detector positions are combined
        into a single observation (Fig.~\ref{fig:cleanspec}), using an
        algorithm which also excludes pixels with anomalous values
        by comparison with a running median.

  \item The spectra are extracted from each observation. Since the output of
        each element is not individually resolved along the slit, the 
        positions must be calculated from known offsets.

  \item Extracted spectra are divided by the extracted {\it fiber 
        flatfield} exposure, to correct for variations in element
        transmission.

  \item The spectra are reformatted into an ($x,y,\lambda$)
        datacube (Fig.~\ref{fig:1cubeim}), using the known mapping
        between position in the field and position in the slit, 
        and by fitting a bidimensional surface to the
        spatial points at each wavelength increment. 

  \item If spatial mosaicing has been used to increase the field of view,
        the observations of each mosaic cycle are combined using
        information derived from the telescope offsets or by centroiding
        on known features.
        The individual datacubes are interpolated onto an
        output datacube with sub-pixel accuracy
        (Fig.~\ref{fig:specmap}).

  \item Finally, the datacube is analysed to provide the required
        astrophysical diagnostics such as the distribution of continuum
        light in
        specified bands, the radial velocity field derived from the
        shift in 
        any of the spectral features present and emission line ratios.

\end{enumerate}

This procedure has been coded into a series of IRAF tasks specially
written for the purpose. 

\subsection{Observations of NGC4151}

At infrared wavelengths it is possible to view the centres of active
galaxies from which visible light is obscured by dust in the torus. A
fiber-lenslet
IFU gives two-dimensional spectroscopy with the full wavelength resolution
and coverage of a conventional long slit. Thus accurate spectral line
strengths and corresponding velocities can be mapped across an object from a
single observation revealing any association with structures seen in
broad-band images or radio observations. Such an ability to form connections
between spectral and physical features has proved crucial in identifying
line excitation mechanisms.

Understanding excitation is important for studying kinematics as well as
energy transfer. Previous work (Hutchings et al.
1998\markcite{Hutchings}, Gallimore {\it et al. } 1997\markcite{Gallimore})
0suggests that the optical narrow line regions (NLR) of Seyfert galaxies are
driven by photoionization via a cone of radiation from the active
nucleus. This scenario is compatible with gravitationally dominated dynamics
in those regions, indicated also by measurements of velocity dispersion
(Nelson \& Whittle 1996)\markcite{Nelson}. However, emission may also occur
through shock excitation (Genzelet al. 1995)\markcite{Genzel}, in
which case corresponding velocity measurements cannot be used to constrain
the overall galactic gas kinematics.  The strongest near-infrared emission
lines in active galaxies are two forbidden [Fe\,II] transitions ($1.26$ and
$1.64 \mu$m), produced in regions where hydrogen is partially
ionized. Usually the transition between ionized and neutral hydrogen is very
sharply defined, so there has been recent speculation as to the how the
partially ionized regions may arise (Veilleux et al.
1997)\markcite{Veilleux}. Proposed mechanisms involve X-ray photoionization
of optically thick narrow line clouds, or shocks from either supernova
remnants in starbursts or the interaction of a radio jet with the
interstellar medium.

The $1.257\mu m$ [Fe\,II] line in the J-band falls close to Pa$\beta$, at
$1.28\mu m$; the ratio [Fe\,II]/Pa$\beta$ has been found to be greater where
emission is due to shock excitation than where photoionization dominates
(Simpson et al. 1996)\markcite{Simpson}, so helps to separate X-ray
illumination from radio jet induced shocks. The two emission lines can be
observed simultaneously with the 150 lines/mm grating in CGS4 (albeit
without much bare continuum for reference), and, being so close together,
provide a combined indicator which is relatively insensitive to reddening.

As an example, we show preliminary results for NGC4151, one of a number of
nearby Seyfert galaxies observed in March 1998.  To search for extended line
emission over a field much larger than the nucleus, we constructed a mosaic
using offsets from the nucleus. This also resulted in better sampling
depending on location within the field.

Fig.~\ref{fig:specmap} shows data from 51 IFU observations in 7 overlapping
fields of view. The observations were combined into a single $78\times
106\times 257$ element datacube as described in Section 6.1.  In this
galaxy, the axes of the radio jet and optical extended narrow line region
are separated in the sky by $\sim 25$ degrees. The intention was to trace
the infrared NLR emission far enough away from the nucleus to see whether it
follows the jet or the optical NLR, thereby constraining the mechanism of
excitation.

Fig.~\ref{fig:ironmap} shows maps in the central region of [FeII] and
Pa$\beta$ narrow-line intensity. There is clear evidence that the [FeII]
emission is extended along the optical ENLR axis and marginal evidence for
an extension in the narrow-component of Pa$\beta$. Further details, will be
presented by Turner et al. (in preparation).

\section{Conclusions}

We have demonstrated the potential of lenslet-coupled fibers in the
non-thermal regime for integral field and multiple-object spectroscopy in
the near infrared. The integral field system gives significant advantages
over fiber-only and lenslet-only systems.  We have demonstrated good
performance and utility for astronomy even with a low-cost prototype system
retrofitted to an existing spectrograph.

The multiobject mode of SMIRFS provides a significant multiplex advantage in
observing efficiency over the longslit spectrographs currently used at these
wavelengths and points the way forward for infrared multi-fiber
systems. Fiber systems offer efficient matching of the detector to a large
field of view and make it easier to reach the high spectral resolution
necessary to reject atmospheric OH emission lines.

At present, the number of fibers is limited only by the length of the CGS4
long slit and the small field of UKIRT (4 arcmin).  With a dedicated
spectrograph the multiplex advantage could be much increased. With more work
the thermal emission from the fiber slit could be reduced further without
impacting the throughput, leading to an increase in signal/noise at longer
wavelengths.

With the integral field mode of SMIRFS, we have demonstrated how to use
fiber-lenslet systems for efficient integral field spectroscopy in the near
infrared. This work has also given us invaluable information for the
construction of the recently-commissioned {\it Thousand Element Integral
Field Unit} (TEIFU) and the fiber-lenslet system for the {\it GEMINI
Multiobject Spectrographs} (GMOS).  Both will include dedicated fields for
background subtraction. Although these systems will initially operate at
visible wavelengths, they can be extended with minor modification to work in
the near-infrared.

With fused silica fibers, we can operate efficiently within the wavelength
range where instrumental thermal background is not a problem ($< 1.8
\mu$m). However, work at longer wavelengths will require a cooled
system. Although fibers are not the preferred technology for fully cryogenic
operation, where image slicers (such as the {\it Advanced Image Slicer};
Content 1997\markcite{content97}) are likely to give much better
performance, fibers may still be usable in cooled systems where full
cryogenic temperatures are not required.

It may even be possible to use fibers in fully cryogenic environments since
preliminary results from tests on cold fibers suggest that both the FRD and,
more unexpectedly, the flexibility, are satisfactory for some types of
unmounted fiber under such conditions (Haynes \& Lee, private
communication).  The retention of flexibility is of considerable importance
to {\it multiple} integral field spectroscopy since it suggests that
fiber-lenslet systems could be deployable under cold conditions (see also
Thatte et al.  1998\markcite{thatte98}).

\acknowledgements

We thank Simon Morris for his help with the scientific programme.  We are
indebted to the staff of the Joint Astronomy Centre in Hawaii for help with
the integration of SMIRFS with UKIRT and CGS4, particularly Tom Kerr and Tom
Geballe.  We also thank Adaptive Optics Associates (Carol Dwyer and Brian
McNeil) for their work on the microlens arrays. This work was largely
supported by a grant from the UK Particle Physics and Astronomy Research
Council.

%
%
\newpage

\begin{table}[tbhp]
\caption{Summary of SMIRFS efficiency in MOS mode}
\begin{center}
\begin{tabular}{lcccc}
\hline \hline
                & J     & H     & K     & Reference \\ 
                &       &       &       & fiber \\ 
\hline   
Fused silica    & 0.51  & 0.62  & 0.57  & 7\\
Prediction      & 0.65  & 0.65  & 0.53  &  \\
\hline 
Zirconium fluoride (with output mask) 
                & 0.39  & 0.41  & 0.42  & 4 \\ 
Zirconium fluoride (with no output mask) 
                & 0.56  & 0.65  & 0.75  & 4 \\ 
Prediction      & 0.55  & 0.62  & 0.69  &  \\ 
\hline
\end{tabular}
\end{center}
\label{tab-moseff}
\end{table}

\begin{table}[tbhp]
\caption{Stars observed in NGC1904 (brightest to faintest)}
\begin{center}
\begin{tabular}{llcc}
\hline \hline
Fiber    & star  & V     & B-V   \\ 
\hline 
8    & A41     & 12.94 & 1.41  \\ 12   & A51   & 13.33 & 1.39  \\ 
7    & A45     & 13.84 & 1.18  \\ 13   & A53   & 13.02 & 1.71  \\ 
3    & A50     & 13.49 & 1.76  \\ 14   & A55   & 14.93 & 1.04  \\ 
5    & A42     & 15.28 & 0.97  \\ 2    & A52   & 14.68 & 1.06  \\ 
9    & A54     & 15.78 & 0.86  \\ 6    & A47   & 15.27 & 0.67  \\ 
11   & A44     & 16.12 & 0.82  \\ 4    & A49   & 16.07 & 0.83  \\ 
10   & SKY     &       &       \\ 1    & SKY   &       &       \\ 
\hline
\end{tabular}
\end{center}
\label{tab-mosobs}
\end{table}

\begin{table}[tbhp]
\caption{Summary of SMIRFS-IFU}
\begin{center}
\begin{tabular}{ll}
\hline \hline
{\it Requirements:}\\
\hline
Number of spatial elements      & 72    \\
Spatial sampling                & 0.6 arcsec \\
Field of view                   & 6 $\times$ 4 arcsec \\
Wavelength range                & 1--1.8$\mu$m optimized\\
                                & 1--2.5$\mu$m total\\
\hline
{\it Design parameters:}\\
\hline
Focal ratio from telescope      & f/36 \\
Input lenslet pitch             & 412$\mu$m \\
Input focal length (air)        & 5.35mm\\
Fiber core diameter             & 150$\mu$m \\
Fiber input focal ratio         & F/13 \\
Fiber output focal ratio        & F/8 \\
Output lenslet pitch            & 793$\mu$m \\
Focal ratio into CGS4           & F/40 \\
\hline
\end{tabular}
\end{center}
\label{tab-ifu}
\end{table}

\begin{table}[tbhp]
\caption{Details of SMIRFS-IFU fibers}
\begin{center}
\begin{tabular}{ll}
\hline \hline
Core material                   & Fused silica low OH\\
Core diameter                   & $151 \pm 2\mu$m \\
Numerical aperture              & $0.22 \pm 0.02$ \\
Core-cladding ratio             & 1.1 \\
Buffer material                 & Polyimide \\
Buffer diameter                 & $195 \pm 3\mu$m \\
Transmission range              & $0.38-2.5\mu$m \\
Fiber length                    & 1.2m \\
\hline
\end{tabular}
\end{center}
\label{tab-fibers}
\end{table}

\begin{table}[tbhp]
\caption{Summary of SMIRFS-IFU efficiency in the J and H bands.}
\begin{center}
\begin{tabular}{lll}
\hline \hline
Loss mechanism                          & J     & H \\
\hline
Misalignment with secondary mirror      & 0.99  & 0.99  \\
Fiber transmission                      & 0.998 & 0.997 \\
Lenslet fresnel loss                    & 0.921 & 0.923 \\
Lenslet + adhesive transmission         & 0.996 & 0.968 \\ 
Input fiber--lenslet geometrical loss   & 0.811 & 0.795 \\ 
Output fiber--lenslet geometrical loss  & 0.969 & 0.964 \\
IFU-spectrograph pupil geometrical loss & 0.940 & 0.948 \\
Slit projection mirrors                 & 0.910 & 0.929 \\
\hline
Theoretical IFU efficiency              & 0.61  & 0.60  \\
\hline
{\bf Measured IFU efficiency}           & {\bf 0.50}  & {\bf 0.46}  \\      
\hline
\end{tabular}
\end{center}
\label{tab-ifseff}
\end{table}

%
%
\newpage

\enlargethispage{20pt}
\plotone{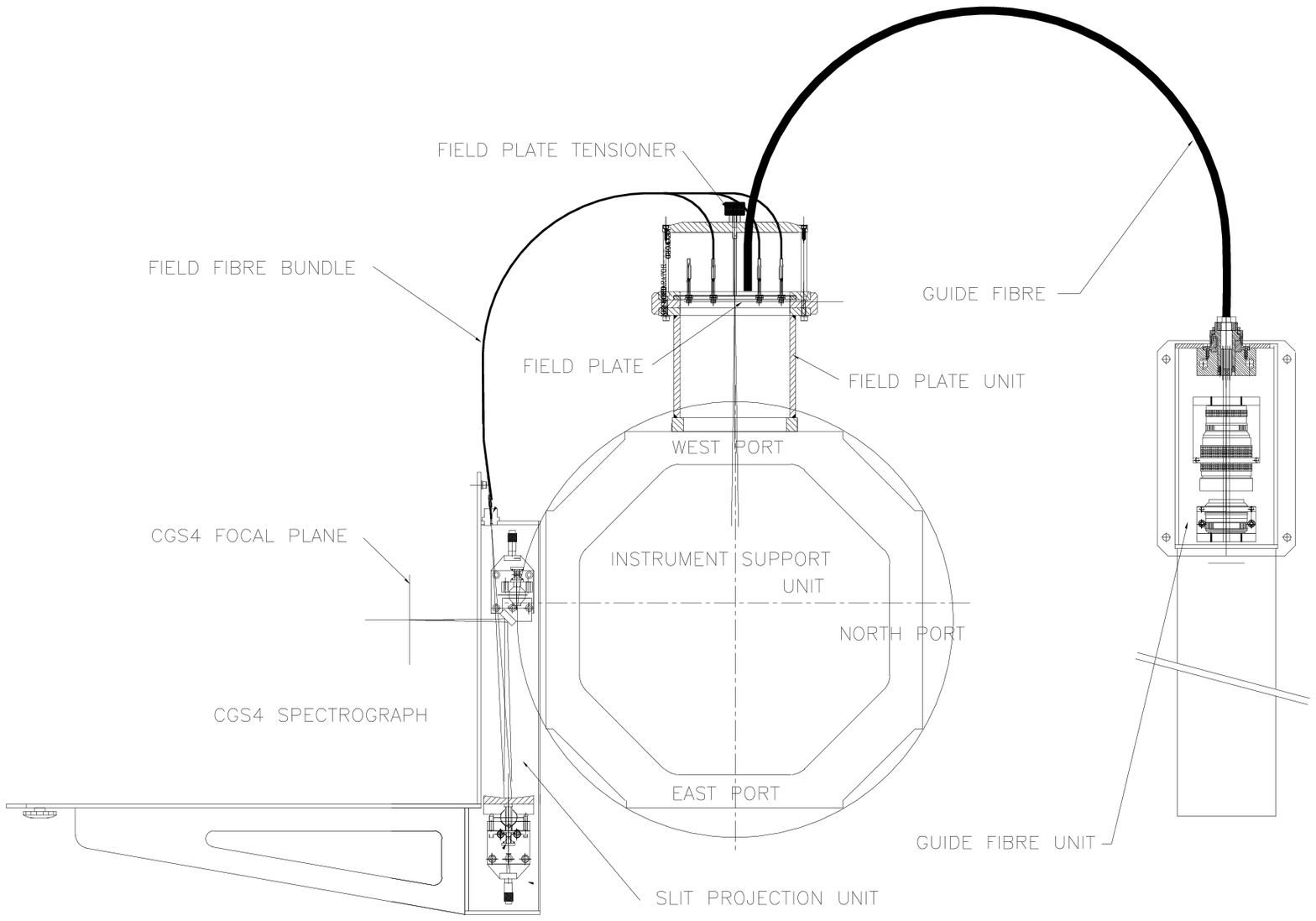}
\figcaption{A plan view schematic of the layout of SMIRFS in
multi-object mode
\label{fig-plan}}

\plotone{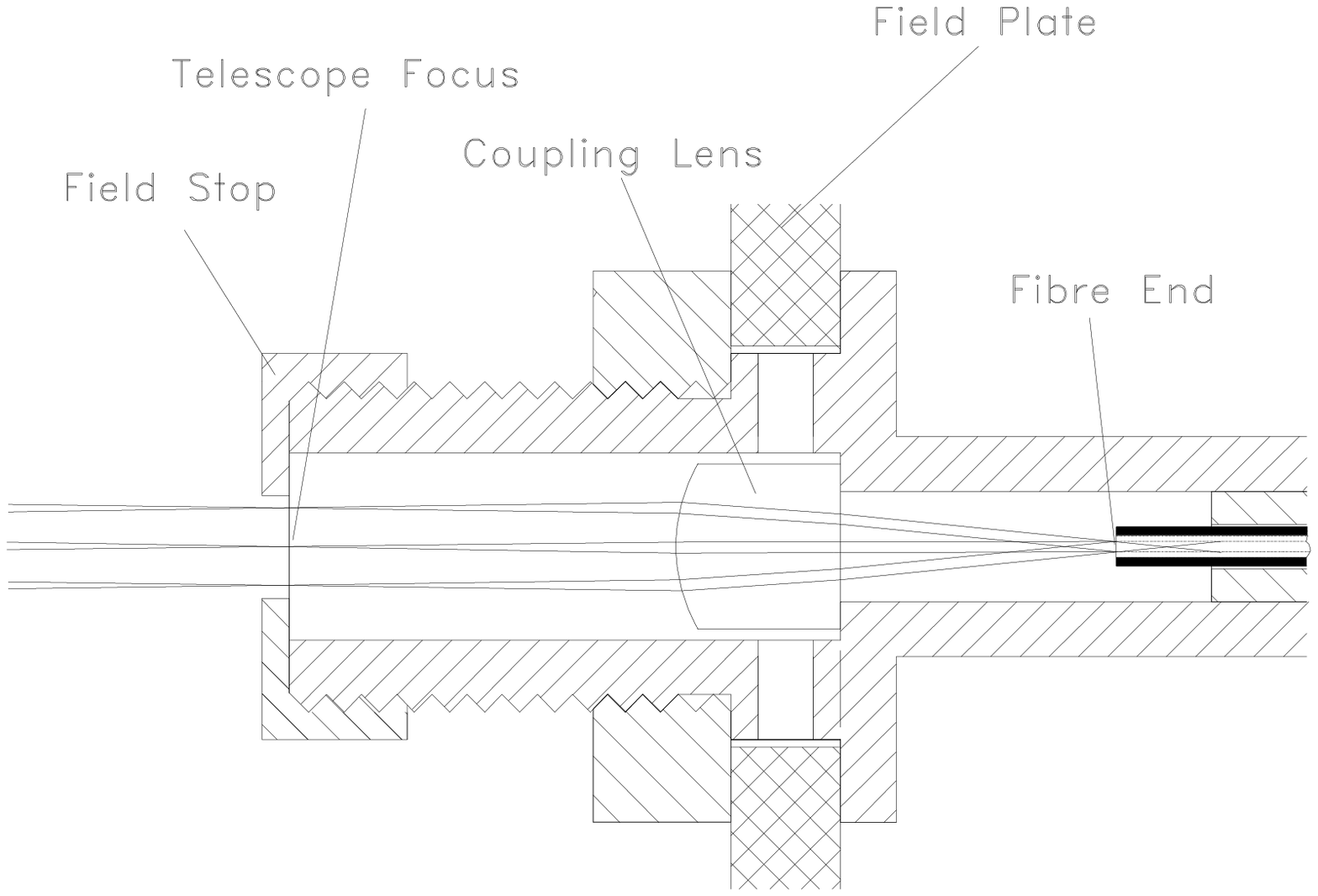}
\figcaption{Schematic of a MOS-mode fiber ferrule showing the lenslet, fiber
and optional field stop.
\label{fig-ferrule}}

\plotone{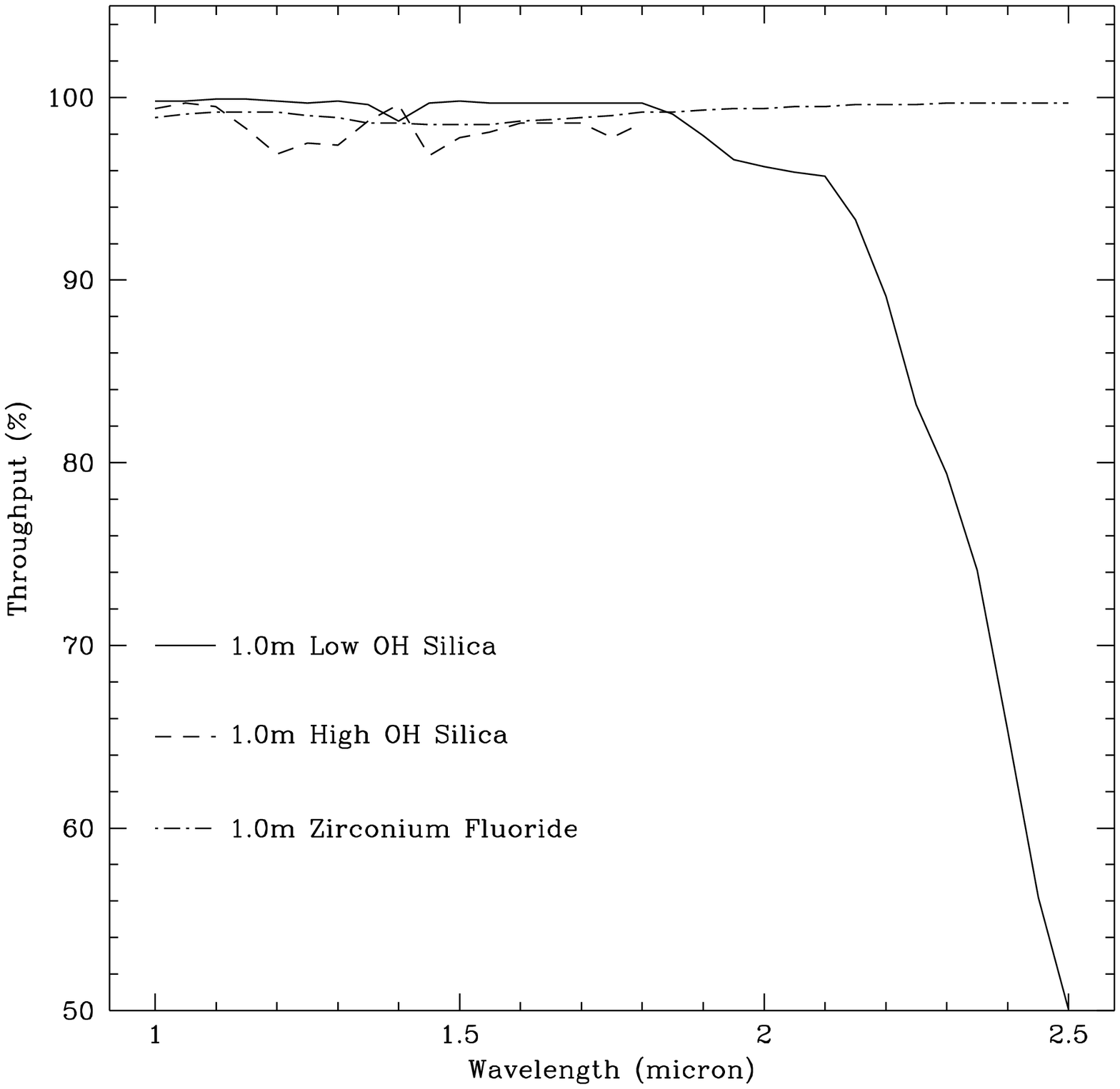}
\figcaption{Transmission of different types of fiber at infrared wavelengths 
for 1m lengths. 
\label{fig-fthro}}

\plotone{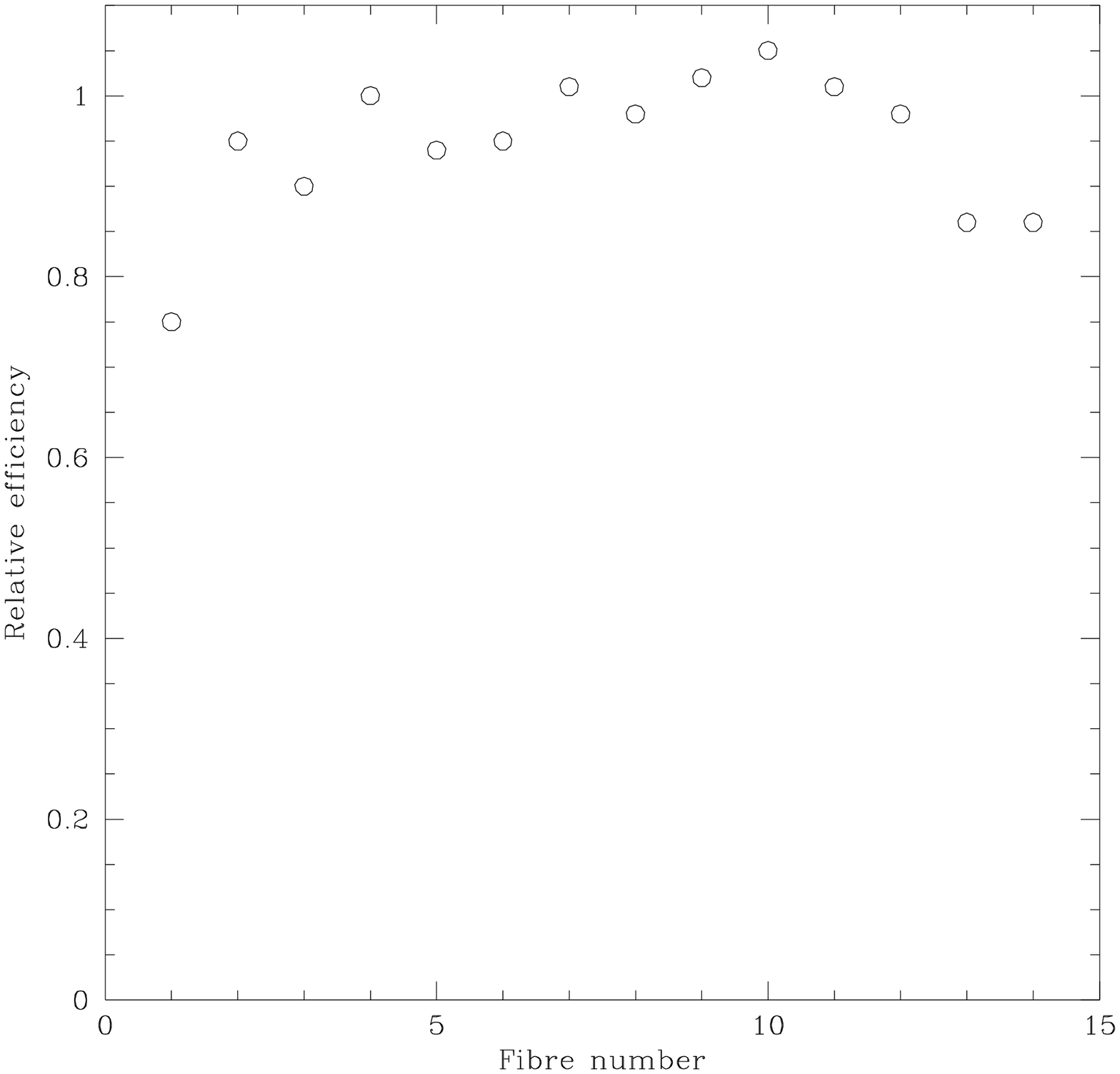}
\figcaption{Example of fiber to fiber variations in throughput in
the MOS mode.
\label{fig-mosflat}}

\plotone{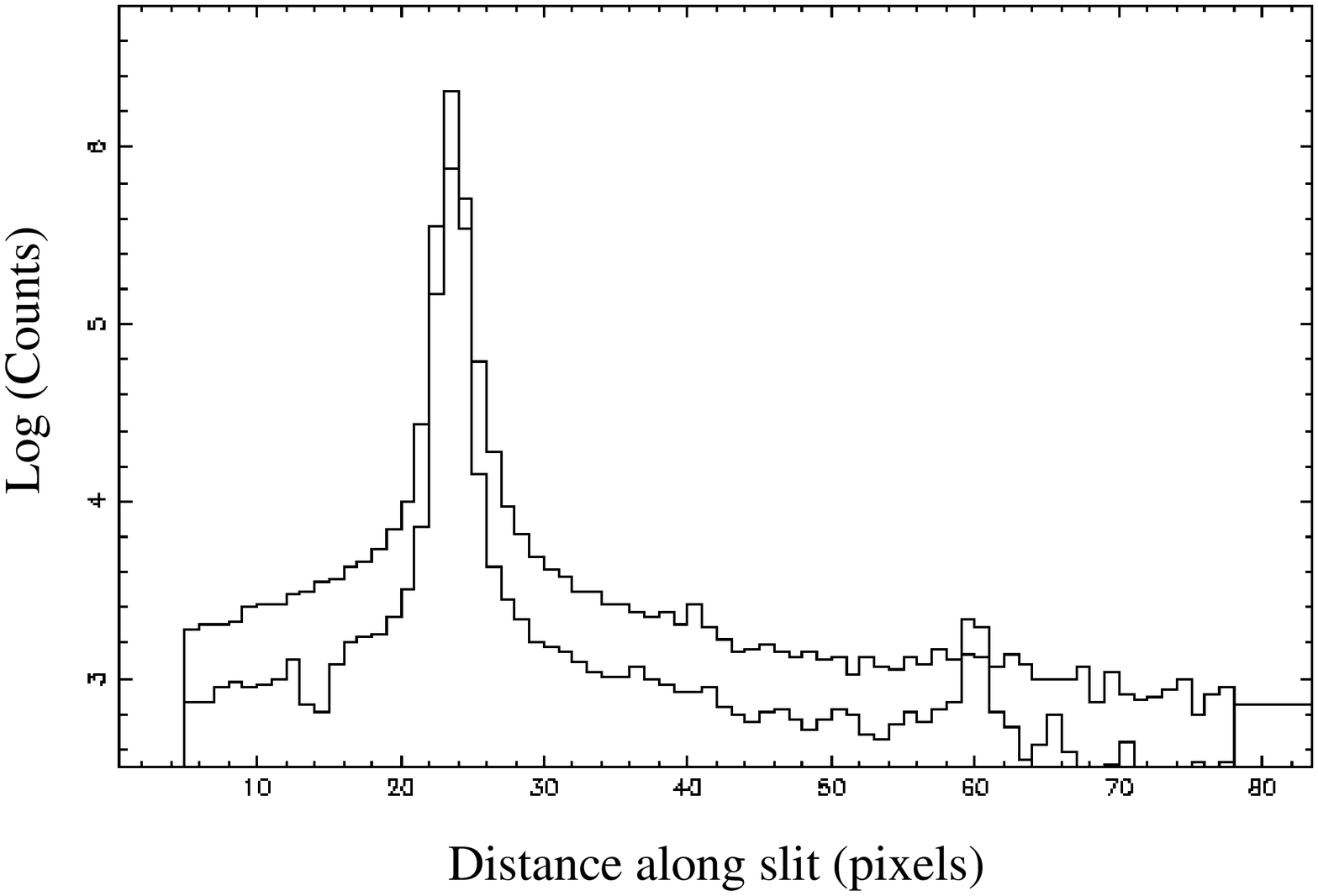} 
\figcaption{Comparison of the spatial profile of a standard star observed
with one fiber of the MOS mode (lower trace) and by CGS4 alone (upper trace)
on a logarithmic scale of intensity. Note that another object has been
caught serendipitously by one of the other fibers.
\label{fig-mospsf}}

\plotone{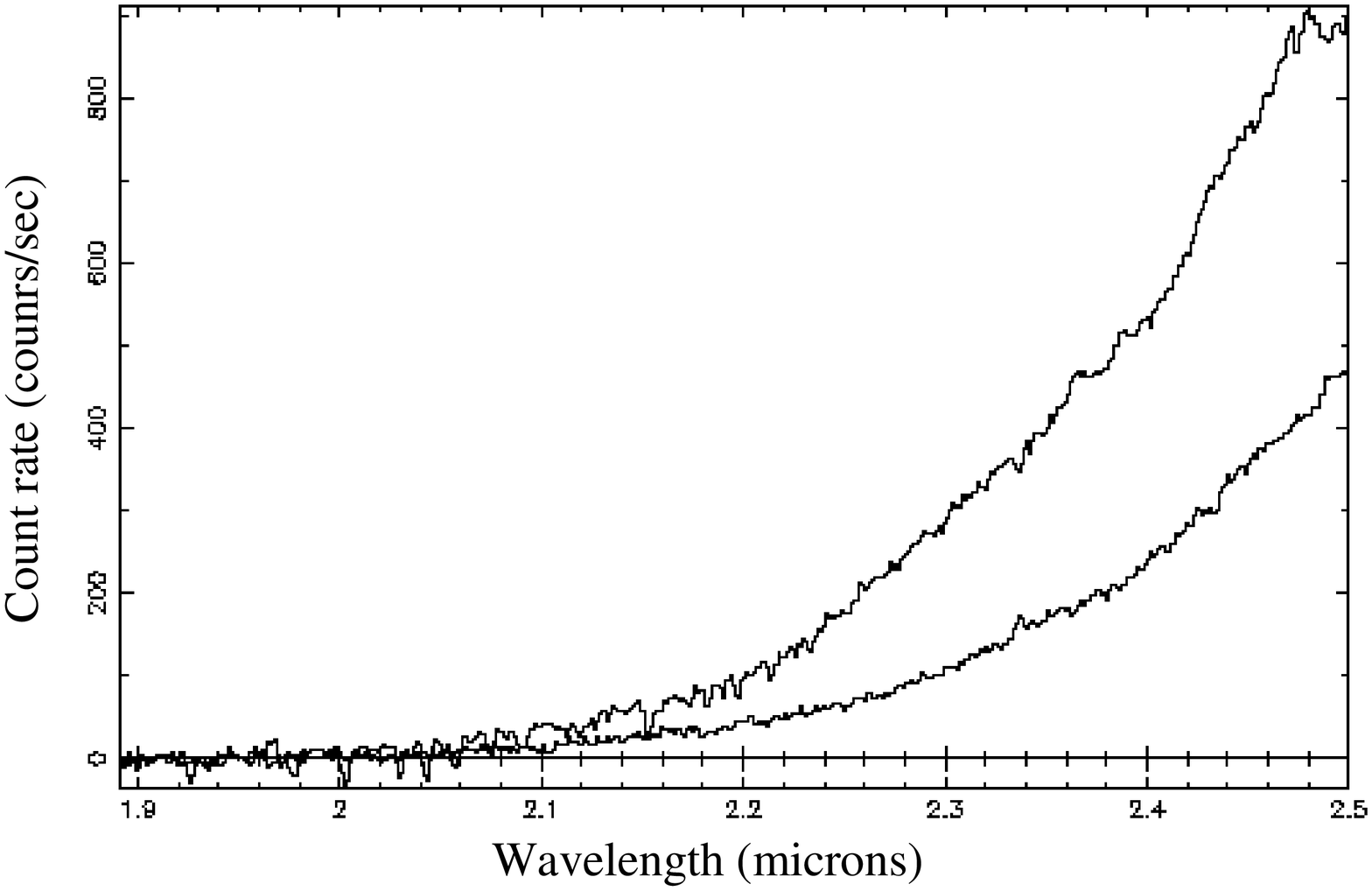}
\figcaption{Background count rate in the K-band for the MOS mode with the
output mask installed for locations within the slit corresponding to the
positions of fibers (upper trace) and between fibers (lower trace).
\label{fig-back}}

\plotone{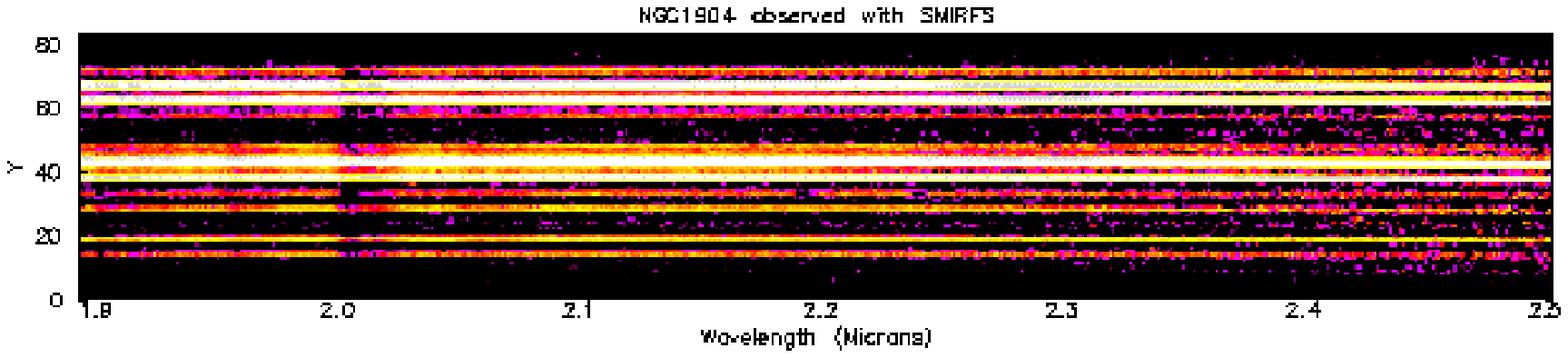} 
\figcaption{MOS-mode background-subtracted
spectrogram for NGC1904. This is a subset of the full detector surface.
\label{fig-image}}

\plotone{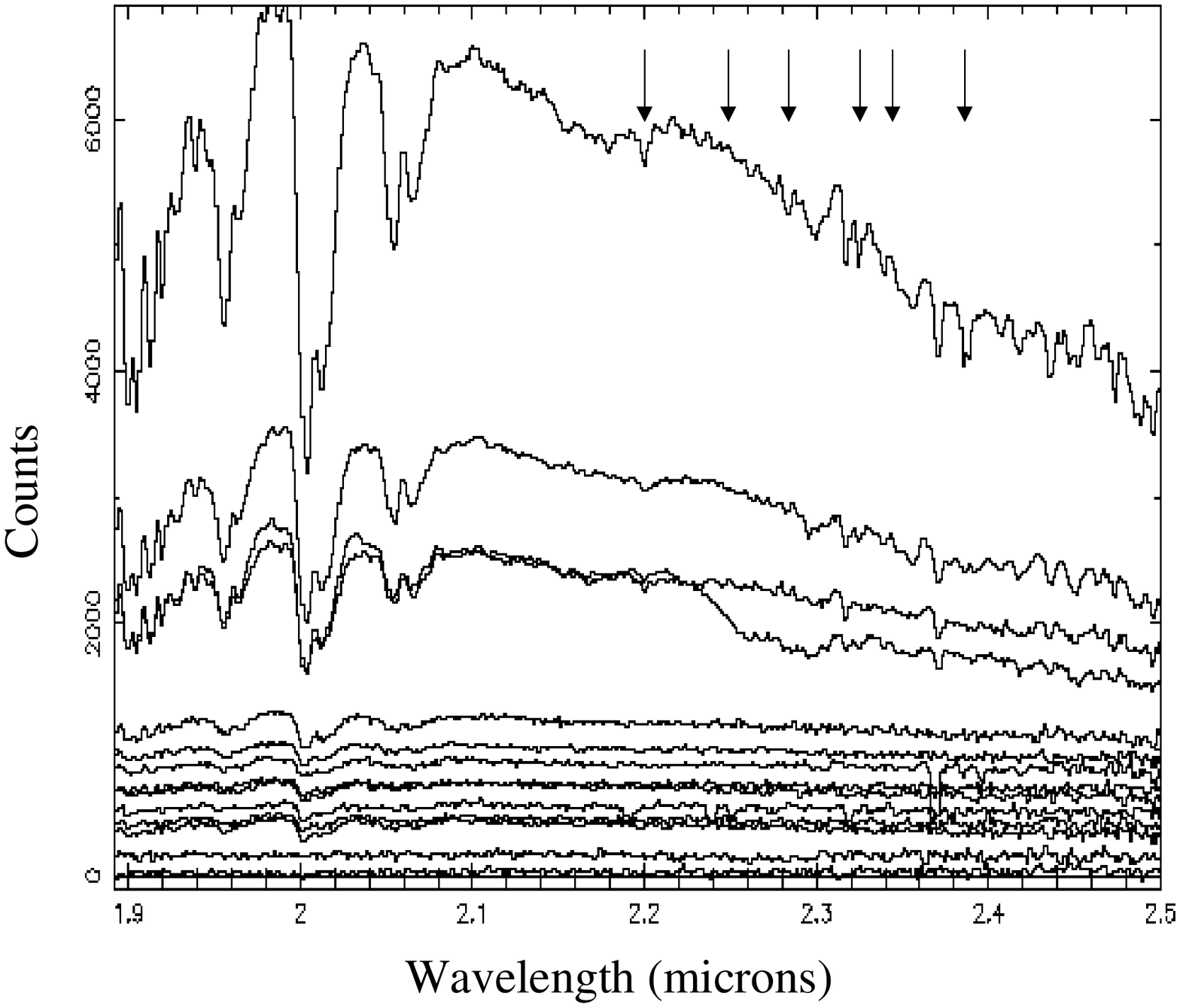} 
\figcaption{Extracted spectra for the NGC1904 observation
(Fig.~\ref{fig-image}). The spectra are offset from each other by 100 counts
for clarity. Arrows mark the location of the NaI and CaI lines mentioned in
the text and some (12)CO lines.
\label{fig-spectra}}

\plotone{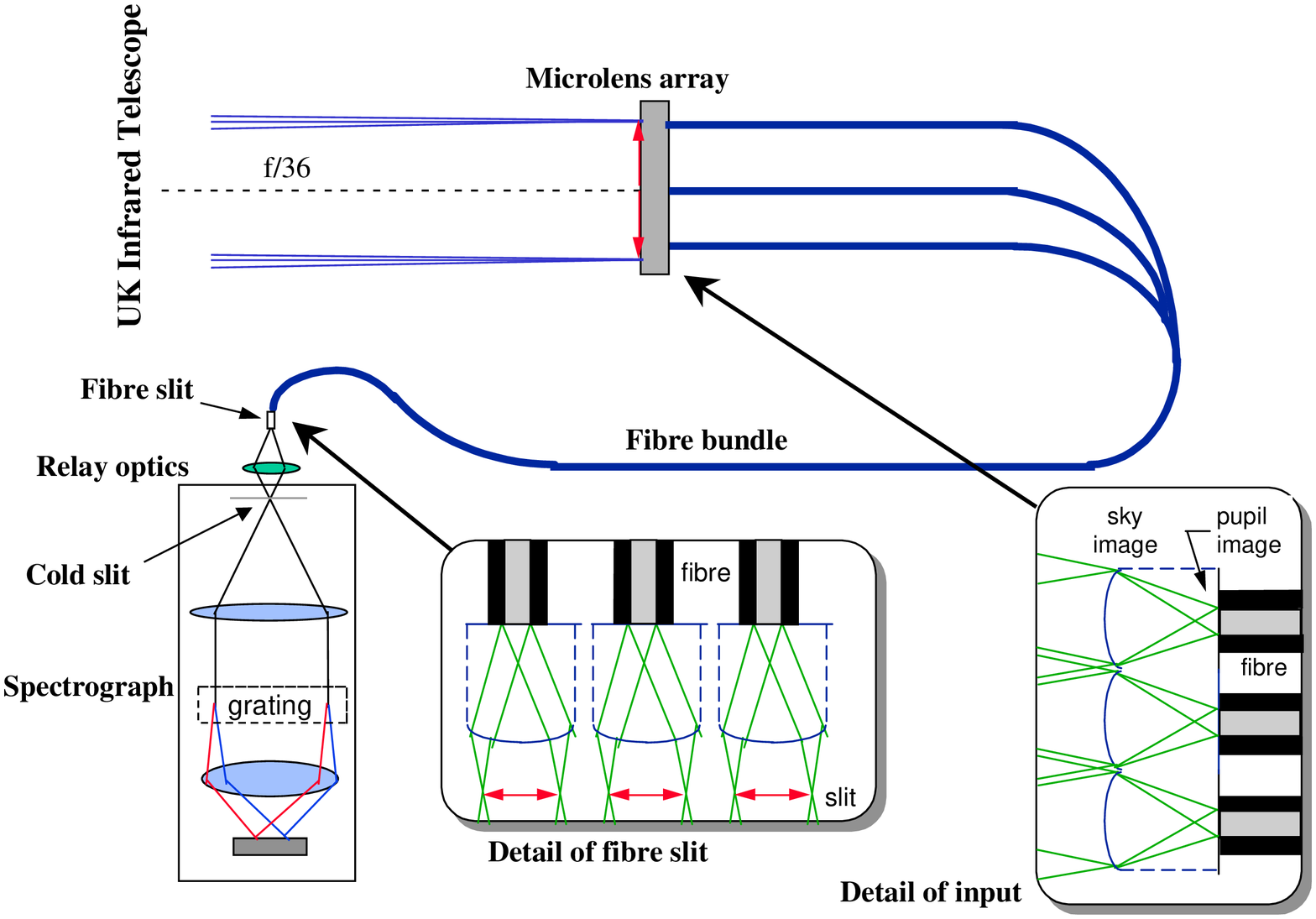}
\figcaption{Basic principle of the integral field unit (IFU).
\label{ifu-scheme}}

\plotone{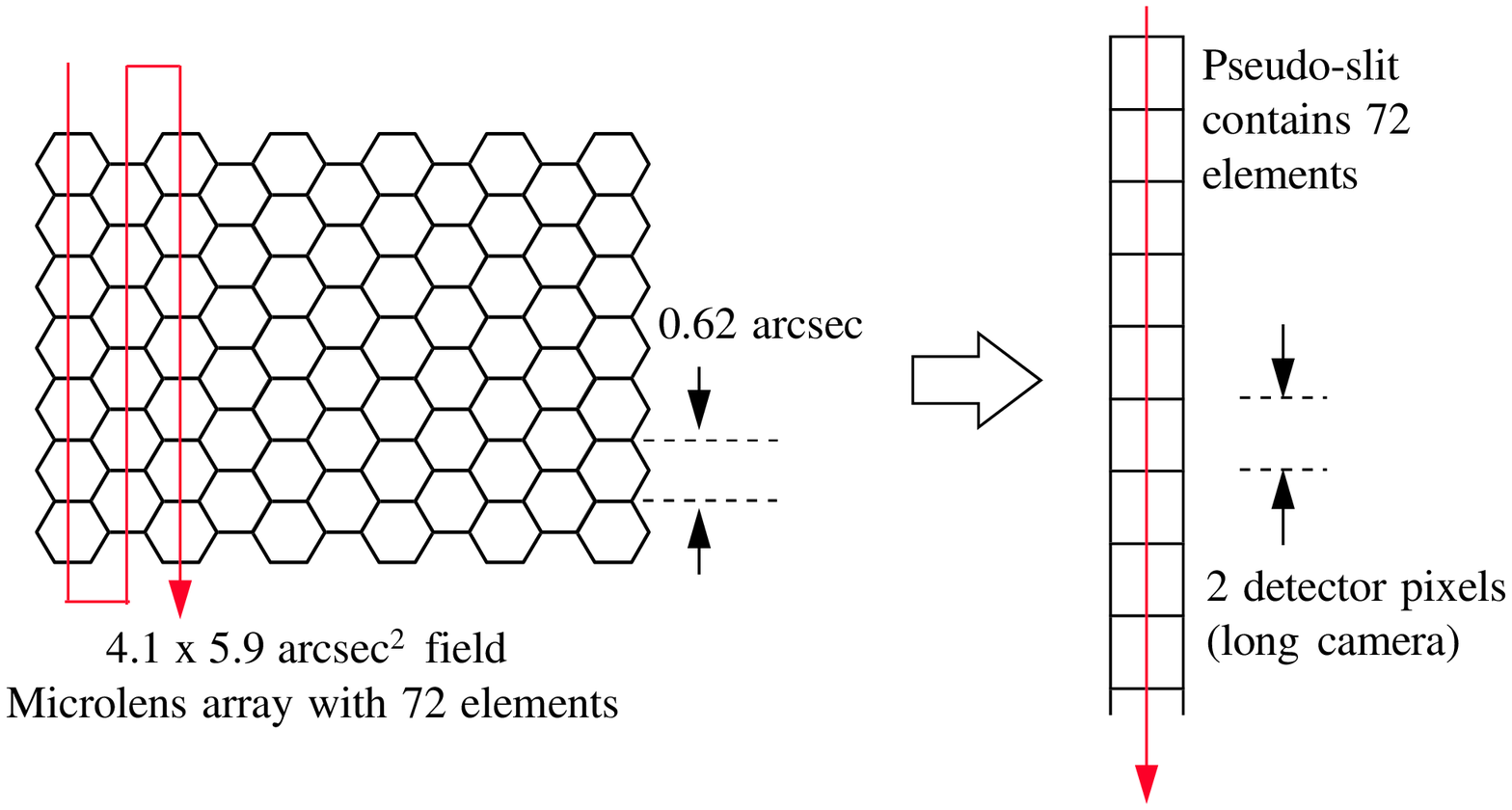}
\figcaption{The field format of the IFU showing how the elements are
reformatted to form the pseudo-slit.
\label{ifu-input}}

\plotone{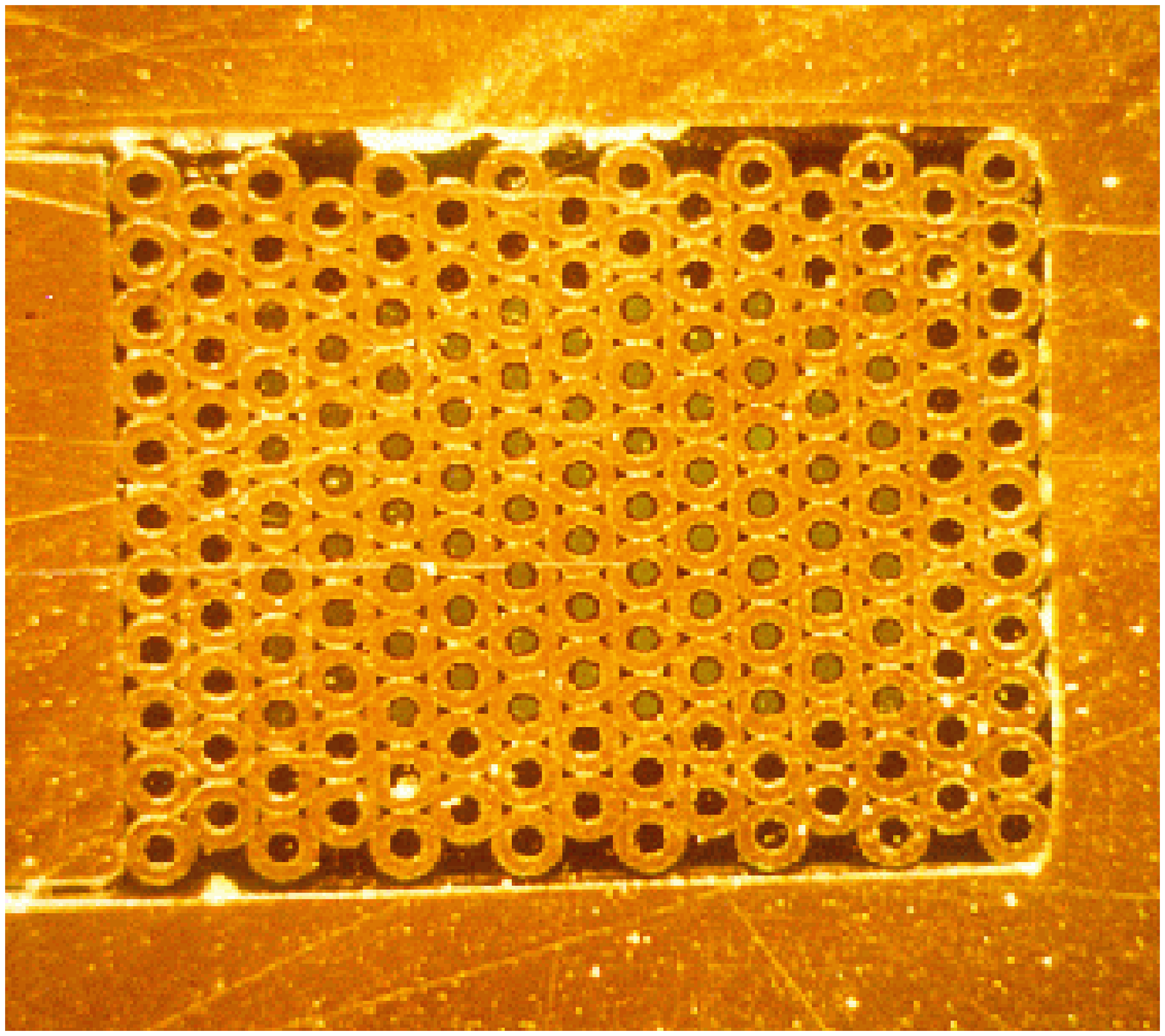} 
\figcaption{The IFU input fiber holder. The fibers are located inside the
microtubes. Some microtubes at the edge of the array are not
populated. The assembly is shown after polishing but prior to the
attachment of the lenslet array. A few large-scale scratches can be seen
where a contaminant has been accidentally introduced into the polishing
medium but these mostly miss the fibers and do not affect the
performance on the system.
\label{ifu-tubes}}

\plotone{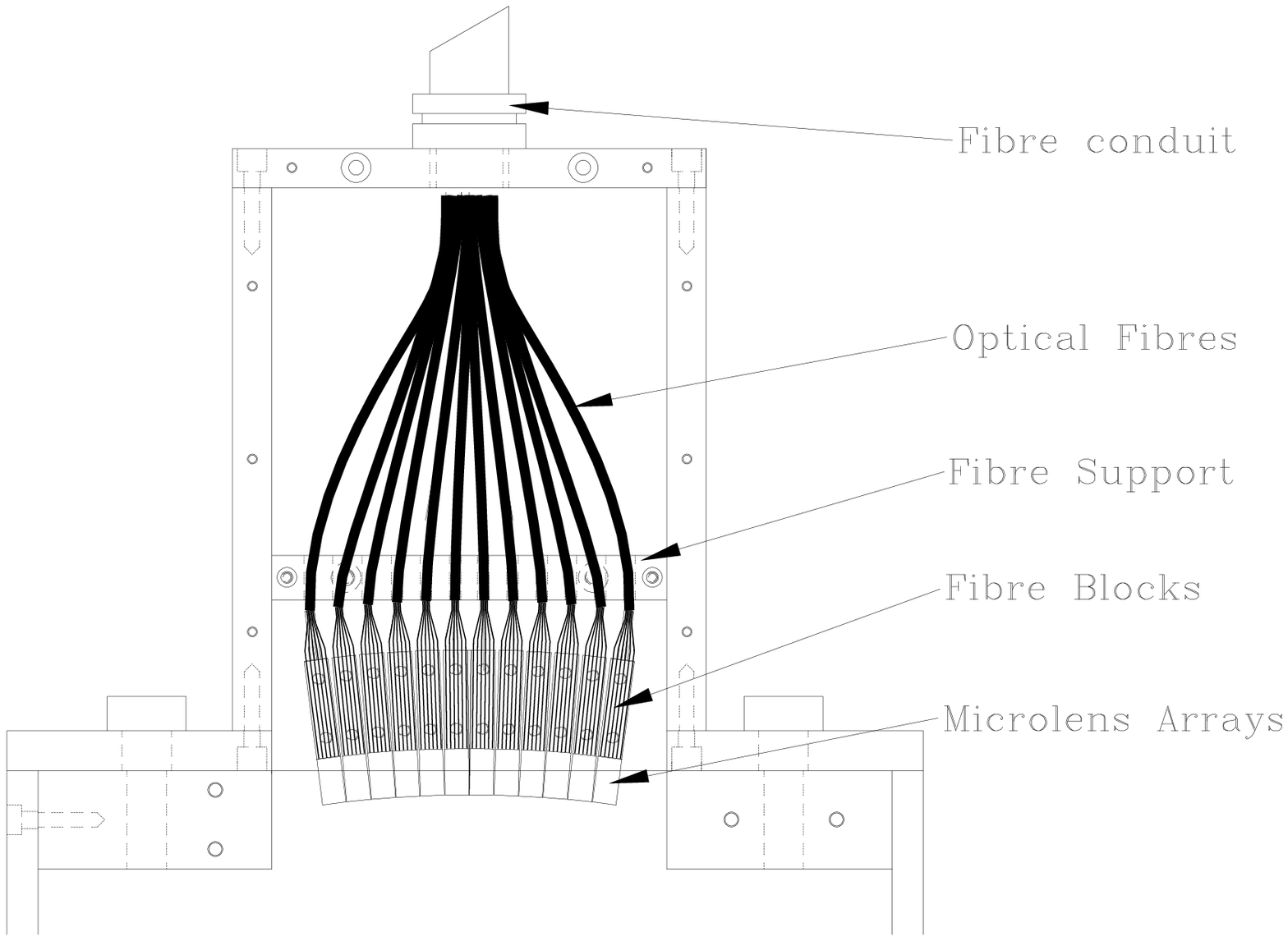}
\figcaption{The IFU pseudo-slit. This shows how the fibers are arranged at
the slit and attach to the lenslet arrays. The curvature of the slit is
required to match the spectrograph pupil.
\label{ifu-slit}}

\plotone{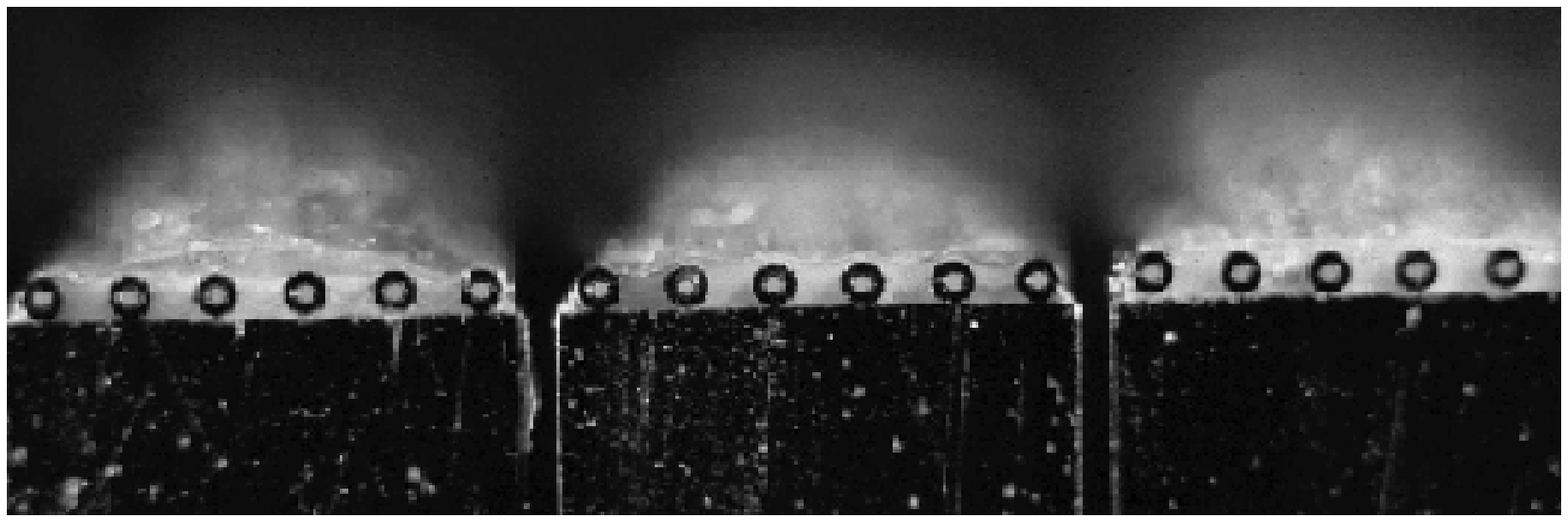}
\figcaption{Part of the IFU slit assembly showing the polished fibers
prior to the attachment of the lenslet arrays.
\label{ifu-slitblock}}

\plotone{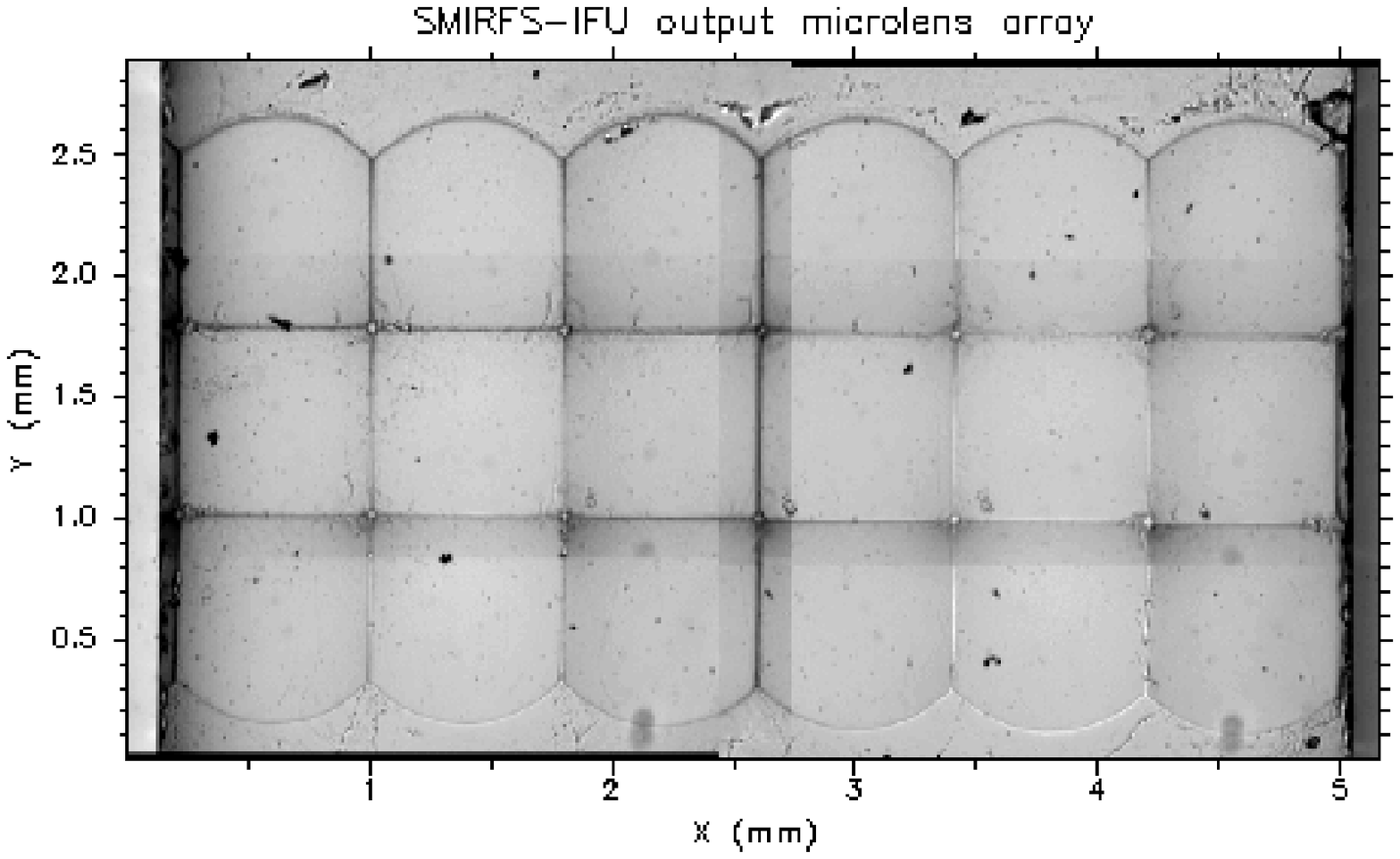}
\figcaption{One IFU output lenslet array. The dark regions are defects
in the lenslets or dust particles. Only the middle row of 6 is used.
\label{ifu-outlens}}

\plotone{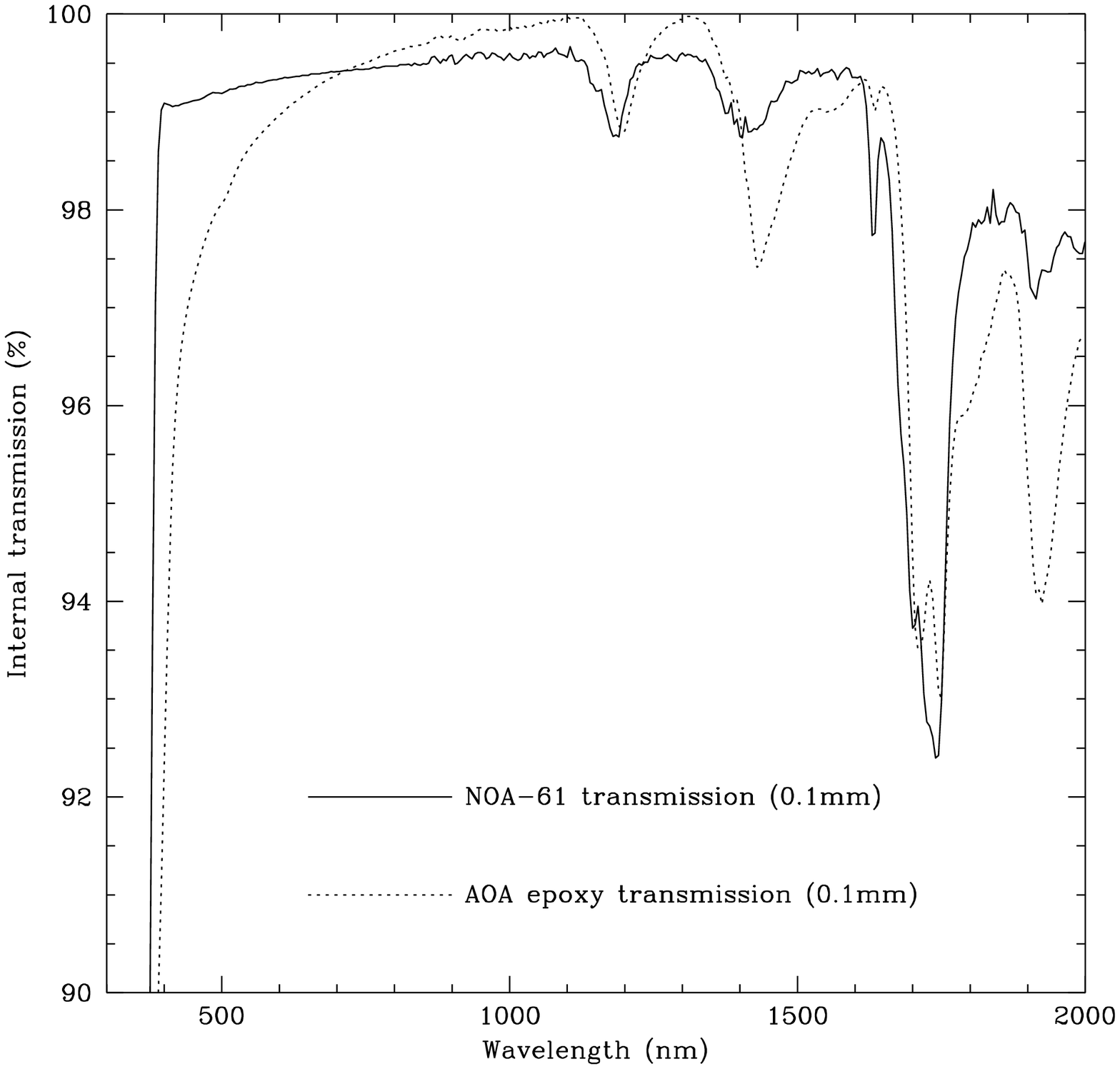} 
\figcaption{Internal transmission of the epoxy used
in the IFU lenslet construction and the adhesive used to bond them to the
fibers. The data are scaled from our measurements to the typical thickness
used in the SMIRFS IFU, as indicated. See Lee (1998) for further details.
\label{fig-epoxyglue}}

\plotone{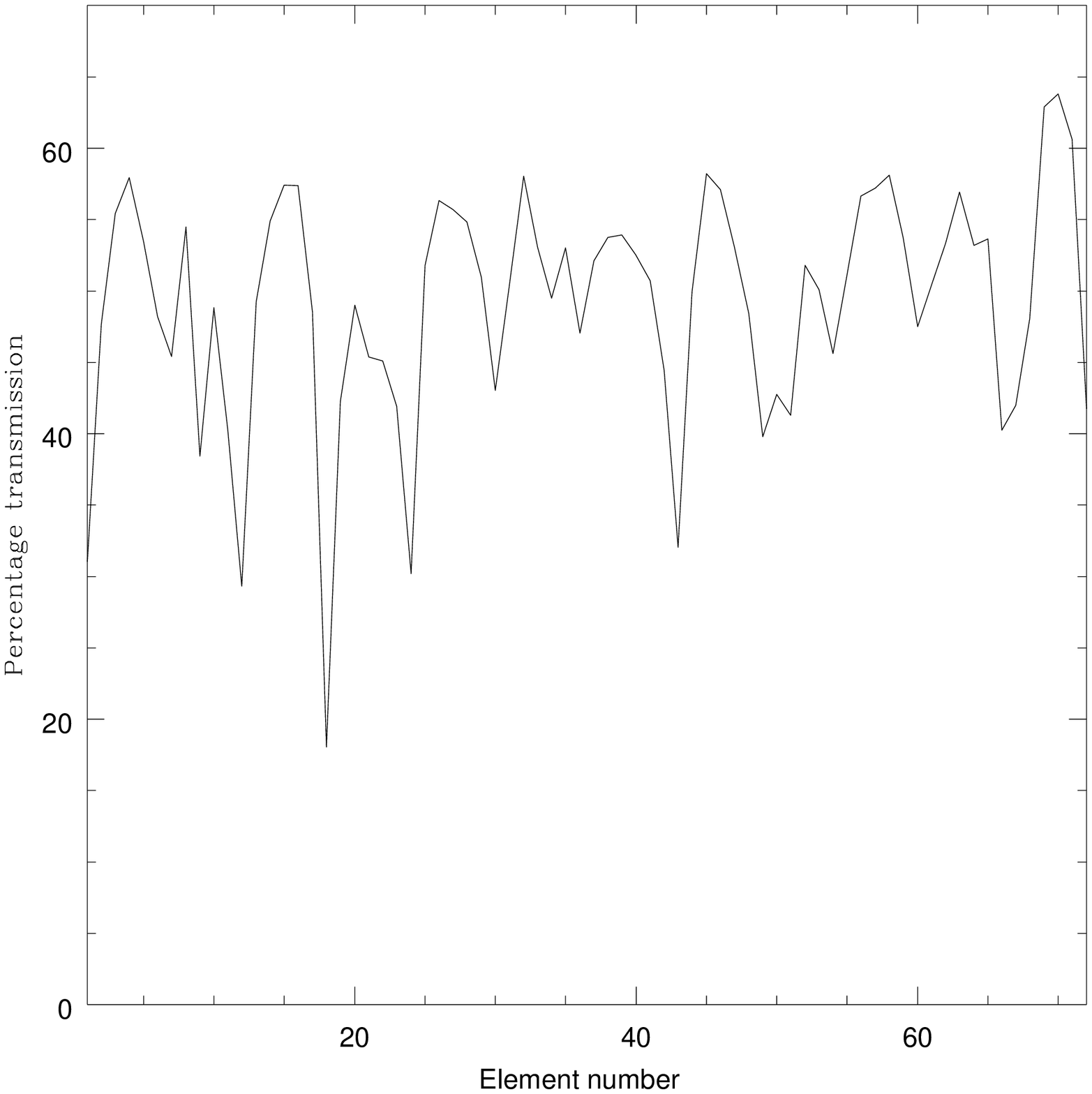} 
\figcaption{Fiber-to-fiber throughput variations for the integral field mode
showing the intensity of light recorded along the slit for uniform incident
radiation integrated over wavelength.  Element 18 has a broken fiber. For
elements 12, 24 and 32, the fibers became partially unbonded from the slit
during polishing.
\label{ifu-flat}}

\plotone{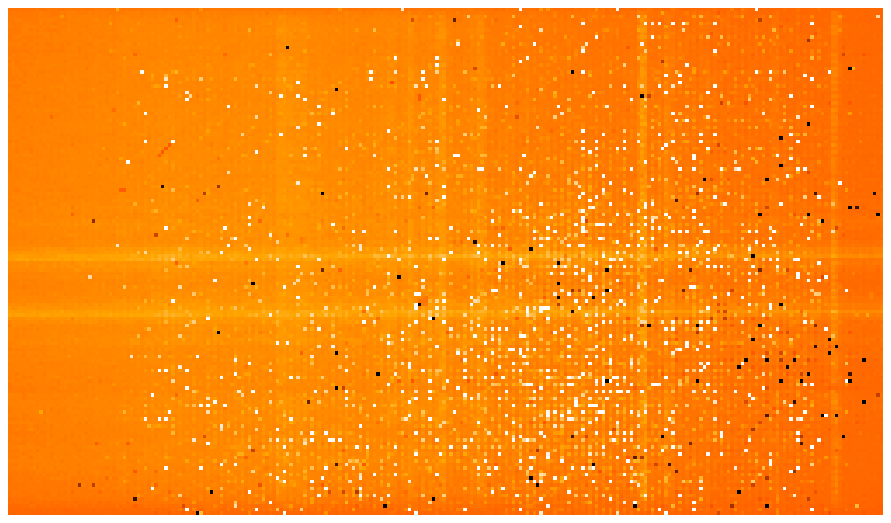}
\figcaption{An example of a raw integration from CGS4. Wavelength
increases  from left to right.
\label{fig:rawspec}}

\enlargethispage{20pt}
\plotone{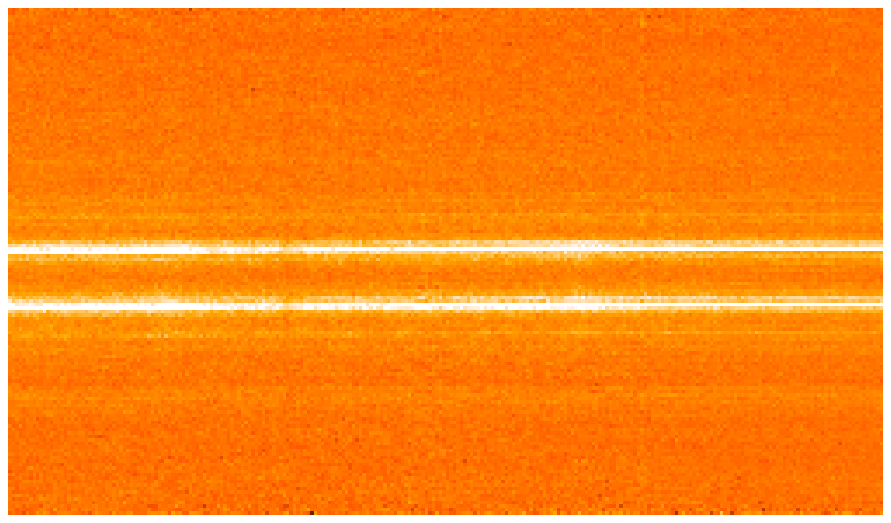} 
\figcaption{An example of a single observation
after sky subtracting, flat fielding and combining integrations.
Wavelength increases  from left to right.
\label{fig:cleanspec}}
 
\plotone{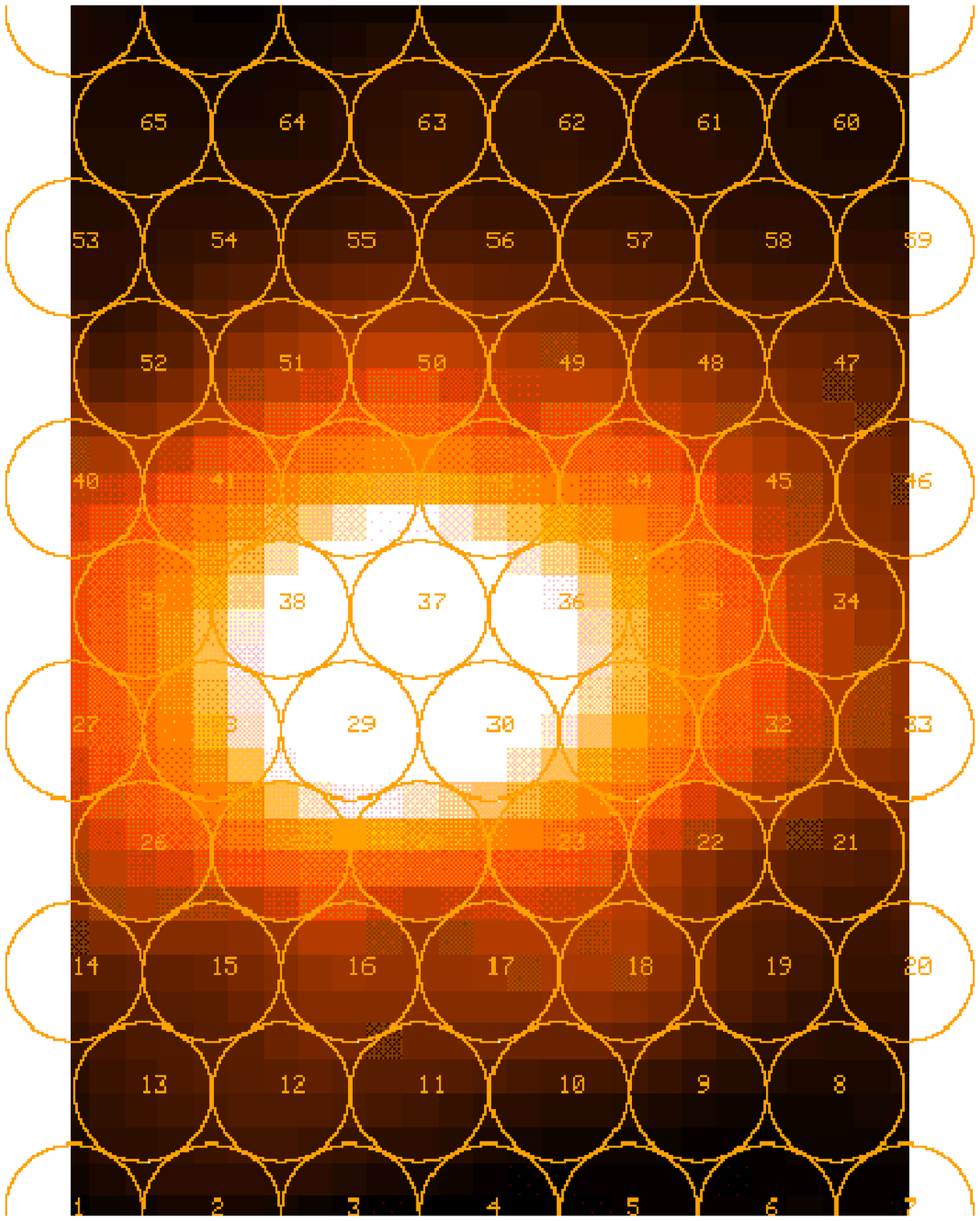} 
\figcaption{An example of a single datacube
collapsed in wavelength, with numbered microlens positions overlaid as circles
\label{fig:1cubeim}}

\plotone{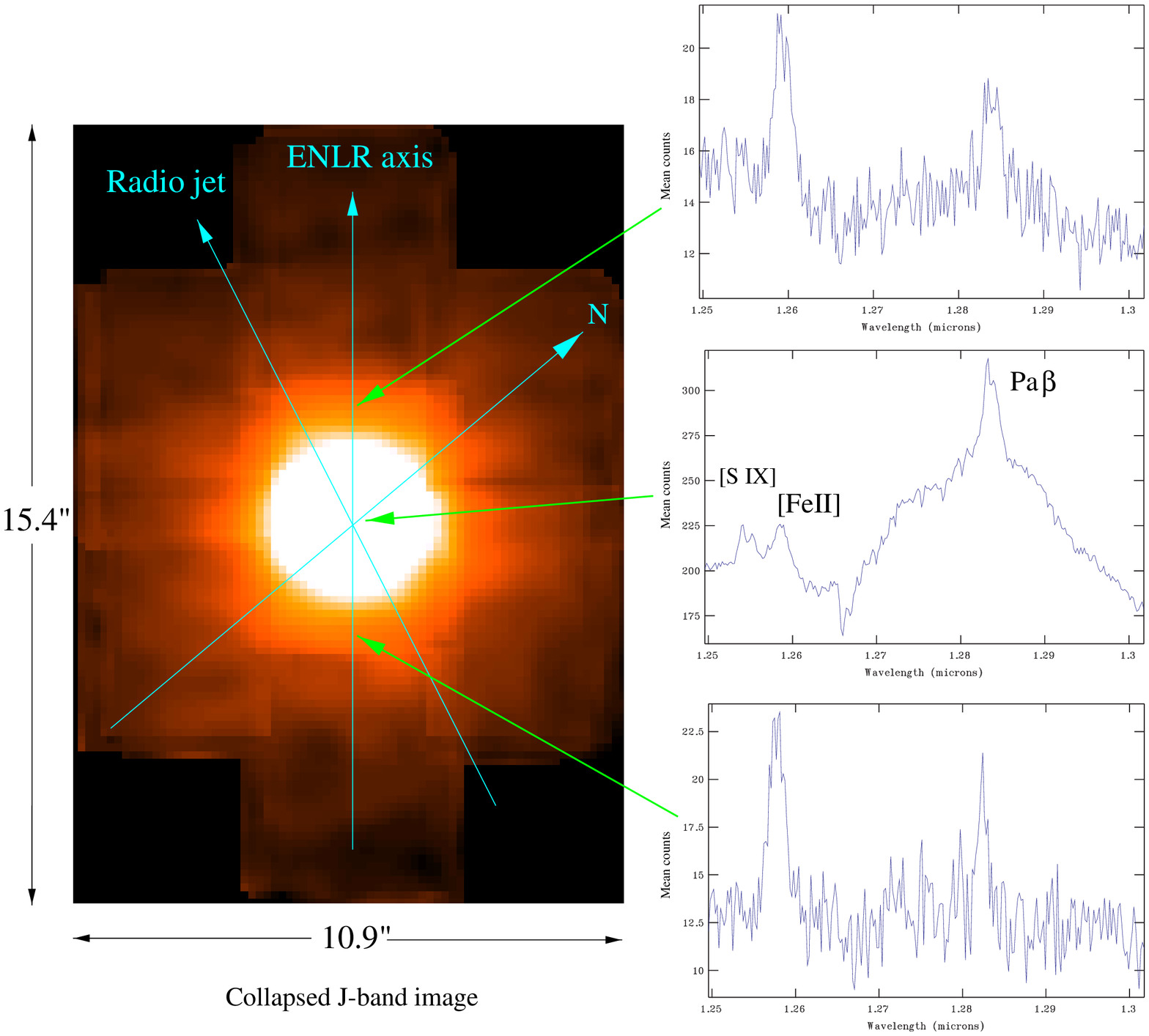} 
\figcaption{Mosaiced integral field mode observations of NGC4151. The image
is a white-light reconstruction from the J-band observations.  The axes of
extended narrow line emission seen at visible wavelengths and the axis of
the radio structure are marked. Examples of spectra along selected lines of
sight are shown. East is left of North.
\label{fig:specmap}}

\plotone{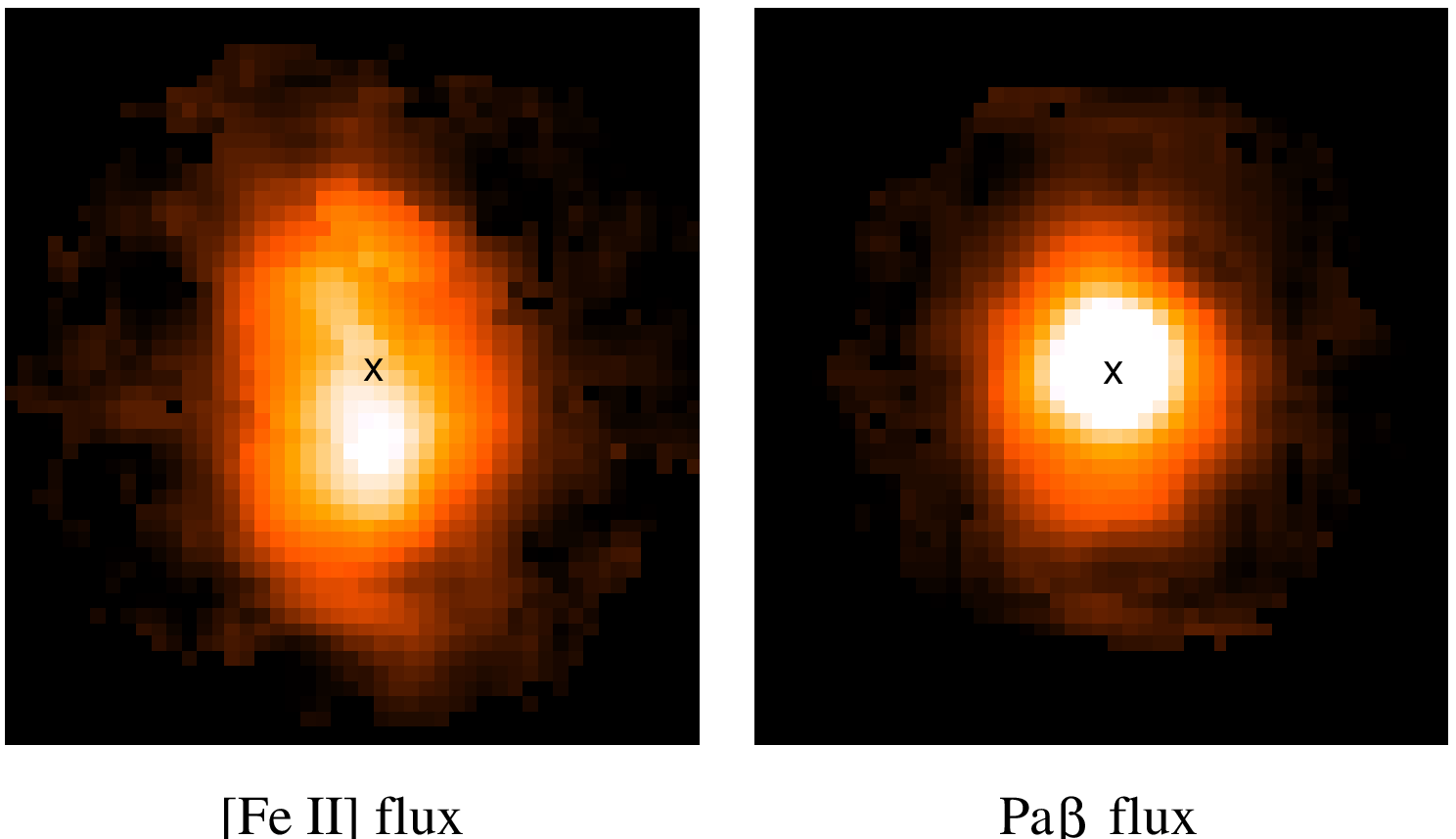}
\figcaption{Continuum-subtracted image of [FeII] emission 
and the narrow components of the Pa$\beta$ emission. Each map covers the
central 8$\times$6 arcsec region shown in Fig.~\ref{fig:specmap}. The
nucleus is marked by a cross.
\label{fig:ironmap}}


\begin{references}

\reference{jras94}
Allington-Smith, J., Breare., J., Ellis, R., Gellatly, D., Glazebrook. K.,
Jorden, P., MacLean, J., Oates, A.P., Shaw, G., Tanvir, N., Taylor, K.,
Taylor, P., Webster, J. \& Worswick, S.P. 1994, PASP 106, 983-991

\reference{jras97}
Allington-Smith, J., Content, R., Haynes, R., \& Lewis, I. 1997,
\procspie, 2871, 1284-1294

\reference{ac98}
Allington-Smith, J., \& Content, R. 1998, PASP, 110, 1216-1234

\reference{bacon95}
Bacon, R., Adam, G., Baranne, A., Courtes, G., Dubet, D., Dubois, J. P.,
Emsellem, E., Ferruit, P., Georgelin, Y., Monnet, G., Pecontal, E., Rousset,
A. \& Say, F.  1995, \aap, Suppl. 113, 347B

\reference{content97}
Content, R. 1997, \procspie, 2871, 1295-1305

\reference{cuby94}
Cuby, J-G. \& Mignoli, M. 1994, \procspie, 2198, 98-109

\reference{Gallimore} 
Gallimore. J., Baum, S., \& O'Dea, C. 1997, Nature, 388, 852

\reference{Genzel}
Genzel, R., Weitzel, L., Tacconi-Garman, L., Blietz, M., Cameron, M.,
Krabbe, A., Lutz, D., Sternberg, A.  1995, ApJ, 444, 129

\reference{haynes95}
Haynes, R. 1995, PhD thesis, University of Durham

\reference{haynesetal95}
Haynes, R., Sharples, R. \& Ennico, K. 1995, Spectrum, 7, 4

\reference{haynes98}
Haynes, R., Doel, P., Content, R., Allington-Smith, J., \& Lee, D. 1998,
\procspie, 3355, 788-809

\reference{herbst95} 
Herbst, T., Pitz, E., \& Reuther, C. 1995, in `Tridimensional Optical
Spectroscopic Methods in Astrophysics', ASP Conf. Ser. 71, p221, eds G.
Comte \& M. Marcelin

\reference{Hutchings} 
Hutchings, J., Baum, S., Weistrop, D., Nelson, C., Kaiser, M., \& Gelderman,
R. 1998, AJ, 116, 634

\reference{kenworthy98}
Kenworthy, M., Parry, I. \& Taylor, K. 1998, \procspie, 3355, 926

\reference{kerr97} 
Kerr, T. 1997, CGS4 Handbook on the {\it Wordwide Web}
http://www.jach.hawaii.edu/UKIRT.new/instruments/cgs4/handbook.html.

\reference{lee98}
Lee, D., Allington-Smith, R., Content, R. \& Haynes, R. 1998,  \procspie,
3355, 810-820.

\reference{lee-thesis}
Lee, D. PhD thesis, University of Durham 1998

\reference{lefevre98}
LeFevre, O., Vettolani., G., Saisse, M., Maccagni, D., Mancini, D.,
Picat, J. \& Mellier, Y. 1998, \procspie, 3355, 8

\reference{murowinski98}
Murowinski, R., Bond. R., Crampton, D., Davidge, T., Fletcher, J.,
Leckie, B., Morbey, C., Roberts., S., Saddlemyer, J., Sebesta, J.,
Stilburn, J., Szeto. K., Allington-Smith, J., Content, R., Davies, R.,
Dodsworth, G., Haynes, R., Robinson, D., Robertson, D., Webster, J.,
Lee, D., Beard, S., Dickinson, C., Kelly, D., Bennet, R., Ellis, M.,
Williams, P. 1998,  \procspie, 3355, 188-194. 

\reference{Nelson} 
Nelson, C., \& Whittle, M. 1996, ApJ, 465, 96

\reference{nussbaum}
Nussbaum, P., Volkel, R., Herzing, H., Eisner, M. \& Haselbeck S. 1997,
Pure \& Applied Optics, 6, 617

\reference{parry94}
Parry, I., Lewis, I., Sharples, R., Dodsworth, G., Webster, J., Gellatly,
D., Jones, L. \& Watson, F. 1994, \procspie, 2198, 125

\reference{ramsay94}
Ramsay-Howatt S. 1994, \procspie, 2198, 467

\reference{Simpson} 
Simpson C., Forbes, D., Baker, A. \& Ward, M. 1996, MNRAS, 283, 777

\reference{taylor97}
Taylor, K. 1997, \procspie, 2871, 145-149

\reference{terndrup91} 
Terndrup, D., Frogel, J. \& Whitford, A. 1991, ApJ, 378, 742

\reference{thatte98} 
Thatte, N., Tecza, M., Eisenhauer, F., Mengel, S., Krabbe, A., Pak, S.,
Genzel, R., Bonaccini, D., Emsellem, E., Rigaut, F., Delabre, B., \& Monnet,
G. 1998,  \procspie, 3353, 704-715

\reference{Veilleux} 
Veilleux, S., Goodrich, R. \& Hill, G. 1997, ApJ, 477, 631

\end{references}
\end{document}